\documentclass[12pt,preprint]{aastex}

\def\gs{\mathrel{\raise0.35ex\hbox{$\scriptstyle >$}\kern-0.6em 
\lower0.40ex\hbox{$\scriptstyle \sim$}}}
\def\ls{\mathrel{\raise0.35ex\hbox{$\scriptstyle <$}\kern-0.6em 
\lower0.40ex\hbox{$\scriptstyle \sim$}}}
\newcommand{\msun}{M_\odot}

\newcommand{\be}{\begin{equation}}
\newcommand{\ee}{\end{equation}}
\newcommand{\bea}{\begin{eqnarray}}
\newcommand{\eea}{\end{eqnarray}}

\begin{document} 

\title{Quantifying Substructure Using Galaxy-Galaxy Lensing in Distant Clusters}

\author{Priyamvada Natarajan$^{1}$, Jean-Paul Kneib$^{2,4}$, Ian
  Smail$^{3}$ \& Richard Ellis$^{4}$}

\affil{$^1$ Department of Astronomy, Yale University, New Haven, CT,
  06511 U.S.A.}
\affil{$^2$ Laboratoire d'Astrophysique de Marseille, France}
\affil{$^3$ Institute for Computational Cosmology,
University of Durham, South Road, Durham DH1 3LE, U.K.}
\affil{$^4$ Department of Astronomy, Caltech, Pasadena, CA 91105
  U.S.A.
}

\begin{abstract}
We present high-resolution mass reconstructions for five massive
cluster-lenses spanning a redshift range from $z = 0.18$--0.57
utilizing archival {\it Hubble Space Telescope} ({\it
HST}) data and applying galaxy-galaxy lensing techniques.  These
detailed mass models were obtained by combining constraints from the
observed strong and weak lensing regimes. Quantifying the local weak
distortions in the shear maps in terms of perturbations induced by
the presence of galaxy halos around individual bright early-type cluster
member galaxies, we estimate the fraction of mass in the central regions of
these clusters that can be associated with small scale mass
clumps. This technique enables us to directly map the substructure in
the mass range $10^{11}$ -- $10^{12.5}\,\msun$ which we associate with
galaxy-scale sub-halos. The determination of the
mass spectrum of substructure in the inner regions of these clusters
is presented. Constraints are thereby obtained on the masses,
mass-to-light ratios and truncation radii for these sub-halos. We find
that the fraction of total cluster mass associated with individual
sub-halos within the inner 0.5h$^{-1}$ -- 0.8h$^{-1}$ Mpc of these
clusters ranges from 10--20\%. Our results have important implications
for the survival and evolution of substructure in high density cluster
cores and are consistent with the theoretical picture of tidal
stripping of galaxy-scale halos in high-density cluster environments
as expected in hierarchical Cold Dark Matter dominated structure
formation scenarios.
\end{abstract}


\section{Introduction}

The detailed mass distribution within clusters and specifically the
fraction of the total cluster mass associated with individual galaxies
has important consequences for the frequency and nature of galaxy
interactions (Merritt 1983; Moore et al.\ 1996; Ghigna et al.\ 1998;
Okamato \& Habe 1999) in clusters. Knowledge of the dynamical history
of clusters enables a deeper understanding of the physical processes
that shape their assembly and evolution.  The discovery of strong
evolution between $z\sim 0.5$ and the present-day in the morphological
(and star-formation) properties of the galaxy populations in clusters
has focused interest on environmental processes which could effect the
gaseous component and dark matter halo of a cluster galaxy (e.g.\ Dressler
et al.\ 1994, 1997; Couch et al.\ 1994, 1998).

The global tidal field of a massive, dense cluster potential well is
expected to be strong enough to truncate the dark matter halo of a
galaxy whose orbit penetrates the cluster core. Therefore, probing the
tidally truncated extents of galaxy halos in clusters can provide
invaluable clues to the dynamically dominant processes in
clusters. For instance, the survival of individual, compact dark halos
associated with cluster galaxies suggests a high probability for
galaxy--galaxy collisions within rich clusters over a Hubble time.
However, since the internal velocity dispersions of cluster galaxies
($\ls 200$\,{\rm km\,s}$^{-1}$) are significantly lower than their
orbital velocities, these interactions are, in general, unlikely to lead
to mergers, suggesting that the encounters of the kind simulated in
the galaxy harassment picture by Moore et al.\ (1996, 1998) are most
frequent. It is likely that tidal stripping in clusters will lead to 
morphological transformations.

Gravitational lensing has emerged as one of the most powerful
techniques to map mass distributions on a range of scales: galaxies,
clusters and beyond. The distortion in the shapes of background
galaxies viewed through fore-ground mass distributions is independent
of the dynamical state of the lens, therefore, unlike other methods
for mass estimation there are fewer biases in lensing mass
determinations. Here, we focus on mapping in detail the mass
distribution inside the inner regions of massive clusters of galaxies
using Hubble Space Telescope (HST) observations. We exploit the
technique of galaxy-galaxy lensing, which was originally proposed
as a method to constrain the masses and spatial extents of
field galaxies (Brainerd, Blandford \& Smail 1996), which we have
since extended and developed to apply inside clusters (Natarajan \&
Kneib 1996; Natarajan et al. 1998; 2002a). We begin by summarizing 
current results of applying galaxy-galaxy lensing to field galaxies
and the outline the constraints obtained on their masses and halo sizes.

Recent work on galaxy-galaxy lensing in the moderate redshift field
has identified a signal associated with massive halos around typical
field galaxies, extending to beyond 100\,kpc\footnote{We adopt
h=H$_o$/100\,{\rm km\,s}$^{-1}$\,Mpc$^{-1}$=0.7 and
$\Omega_{\Lambda}=0.7$, and scale other published results to this
choice of parameters.} (e.g.\ Brainerd, Blandford \& Smail 1996;
Ebbels et al.\ 2000; Hudson et al.\ 1998; Hoekstra et al.\ 2004). In
particular, Hoekstra et al.\ (2004) report the detection of finite
truncation radii via weak lensing by galaxies based on 45.5 deg$^2$ of
imaging data from the Red-Sequence Cluster Survey. Using a truncated
isothermal sphere to model the mass in galaxy halos, they find a best-fit
central velocity dispersion for an $L^*$ galaxy of $\sigma = 136 \pm 5$
kms$^{-1}$ (68\% confidence limits) and a truncation radius of $185
\pm 30$ kpc (for $h = 0.7$ and $\Omega_{\Lambda} = 0.7$). Similar
analysis of galaxies in the cores of rich clusters suggests that the
average mass-to-light ratio and spatial extents of the dark matter
halos associated with morphologically-classified early-type galaxies
in these regions may differ from those of comparable luminosity field
galaxies (Natarajan et al.\ 1998, 2002a). We find that at a given
luminosity, galaxies in clusters have more compact halo sizes 
and lower masses (by a factor of 2--5) compared
to their field counter-parts. The mass-to-light ratios inferred for
cluster galaxies in the V-band are also lower than that of comparable
luminosity field galaxies.  This is a strong indication of
the effect of the dense environment on the properties of dark matter
halos. It is likely that tidal stripping inside clusters might lead to
morphological transformations as well.

In this paper, we present the determination of the mass function of
substructure in clusters using galaxy-galaxy lensing techniques. A
high resolution mass model tightly constrained by strong and weak
lensing observations is constructed including individual cluster
galaxies and their associated dark matter halos.  We show that over a
limited mass range we can successfully construct the mass function of
sub-halos inside clusters. At the moment only theoretical
determinations are available from high resolution cosmological N-body
simulations. Accordingly the sub-halo mass function is an important
prediction of hierarchical Cold Dark Matter structure formation
models. This innovative application of gravitational lensing enables
computation of the sub-halo mass function directly from observational
data.

The strength of the lensing analysis presented here is the combination
of observed constraints from both strong and weak lensing features
which are used to construct a high resolution mass map of a galaxy
cluster. Anisotropies in the shear field (i.e.\ departure from the
coherent tangential signal) in the vicinity of bright, early-type
cluster members are attributed to the presence of these local
potential wells. Statistically stacking this signal provides a way to
quantify the masses associated with individual galaxy halos. This is
accomplished using a maximum likelihood estimator to retrieve
characteristic properties for a typical sub-halo in the cluster.

The outline of this paper is as follows: in \S2, we describe the
formalism for analyzing galaxy-galaxy lensing in clusters including a
synopsis of the adopted models, \S3 discusses the properties of the
clusters analyzed here. We present the best-fit lens models in \S4 
and discuss the sources of noise in \S5. The results on sub-halo
properties, in particular the sub-halo mass spectrum is presented
in detail in \S6. In \S7 we discuss the future prospects of this
technique and present the conclusions of our work.
\section{Galaxy-galaxy lensing in clusters}

\subsection{Framework for analysis}

For the purpose of extracting the properties of the sub-halo
population in clusters, a range of mass scales is modeled
parametrically. The X-ray surface brightness maps of these clusters
suggests the presence of a smooth, dominant, large scale mass
component. Clusters are therefore modeled as a super-position of a
smooth large-scale potential and smaller scale potentials that are
associated with bright early-type cluster members:
\begin{equation}
\phi_{\rm tot} = \phi_{\rm c} + \Sigma_i \,\phi_{\rm p_i},
\end{equation}
where $\phi_{\rm clus}$ is the potential of the smooth component and
$\phi_{\rm p_i}$ are the potentials of the perturbers (galaxy sub-halos).
The corresponding deflection angle $\alpha_I$ and the amplification
matrix $A^{-1}$ can also be decomposed into contributions from the 
main clump and perturbers,
\begin{eqnarray}
\alpha_I\,=\,{{\mathbf \nabla}\phi_{\rm c}}\,+\,\Sigma_i \,
{{\mathbf \nabla}\phi_{\rm p_i}},\,\,\,A^{-1}\,=\,I\,-\,{{\mathbf 
\nabla\nabla} {\phi_{\rm c}}}\,-\,\Sigma_i \,{{\mathbf \nabla\nabla} 
{\phi_{\rm p_i}}}.
\end{eqnarray}
Defining the generic symmetry matrix,
\begin{displaymath}
J_{2\theta}\,=\, \left(\begin{array}{lr}
\cos {2\theta}&\sin {2\theta}\\
\sin {2\theta}&-\cos {2\theta}\\
\end{array}\right)
\end{displaymath}
we can decompose the amplification matrix above as a direct and linear
sum: \bea 
A^{-1}\,=\,(1\,-\,\kappa_{\rm c}\,-\,\Sigma_i \kappa_{\rm
p})\,I - \gamma_{\rm c}J_{2\theta_{\rm c}} - \Sigma_i \,\gamma_{\rm
p_i}J_{2\theta_{\rm p_i}}, 
\eea 
where $\kappa$ is the magnification
and $\gamma$ the shear.  The shear $\gamma$ is in fact a complex
number and is used to define the quantity $\overline{g}$ the reduced shear 
which is determined directly from observations of the shapes of background 
galaxies: 
\bea
\overline{g_{tot}} = {\overline{\gamma} \over 1-\kappa} =
{{\overline\gamma_c} + \Sigma_i \,{\overline\gamma_{p_i}} \over
1-\kappa_c -\Sigma_i \,\kappa_{p_i}},
\eea 
which simplifies in the coordinate system defined with respect to the
perturber $j$ to (neglecting effect of perturber $i$ if
$i \neq j$): 
\bea 
{\overline g_{tot}}|_j} = { {{\overline \gamma_c}
+{\overline \gamma_{p_j}} \over {1-\kappa_c -\kappa_{p_j }}}, 
\eea
where $\overline g_{tot}|_j$ is the total complex shear induced by the
smooth cluster component and the potentials of the perturbers.
Restricting our analysis to the weak regime, and thereby retaining
only the first order terms from the lensing equation for the shape
parameters (e.g. Kneib et al.\ 1996) we have: 

\be 
{\overline g_I}=
{\overline g_S}+{\overline g_{tot}}, 
\ee 
where ${\overline g_I}=\frac{(a-b)}{(a+b)}\,e^{2i\theta}$\footnote{The
measured shape and orientation are used to construct a complex number
whose magnitude is given in terms of the semi-major axis (a) and
semi-minor axis (b) of the image and the orientation is the phase of
the complex number.} is the distortion 
of the image, ${\overline g_S}$ the intrinsic shape of the source, 
${\overline g_{\rm tot}}$
is the distortion induced by the lensing potentials (eqn.~4).

In the vicinity of perturber $j$ which is then the dominant mass contribution:
\be
\kappa_{p_j} >> \kappa_{p_i} {\rm for } i \ne j
\ee
thus,
\be
{\overline g_I}= {\overline g_S}+{\overline g_{tot}}|_j
= {\overline g_S} +
{ {\overline \gamma_c} \over 1-\kappa_c -\kappa_{p_j}} +
{ {\overline \gamma_{p_j}} \over 1-\kappa_c - \kappa_{p_j}}.
\ee
In the local frame of reference of the perturbers, the mean
value of the quantity ${\overline g_I}$ and its dispersion can be 
computed in circular annuli (at radius $r$ from the perturber center)
{\underline{strictly in the weak-regime}},
assuming a constant value $\gamma_c e^{i\theta_{c}}$ for the smooth 
cluster component over the area of integration. In the frame of
the perturber, the averaging procedure allows efficient subtraction of
the large-scale component, enabling the
extraction of the shear component induced in the background galaxies
only by the local perturber. The background galaxies are assumed to have
intrinsic ellipticities drawn from a known distribution (see the next section 
for further details). Schematically the effect of the cluster
on the intrinsic ellipticity distribution of background sources is to 
cause a coherent displacement ${\tau}$ and the presence of perturbers 
merely adds small-scale noise to the observed ellipticity distribution.

The feasibility of signal detection can be estimated by computing the
dispersion in the shear and hence the signal-to-noise ratio. Averaging
eqn.\ (7) in cartesian coordinates (averaging out the contribution of
the perturbers): 
\bea 
\nonumber \left<{\overline g_I}\right>_{xy} &=&
\left<{\overline g_S}\right> + \left<{\gamma_c e^{i\theta_{c}} \over
1-\kappa_c -\kappa_{p_j}}\right> + \left<{{\overline \gamma_{p_j}}
\over 1-\kappa_c - \kappa_{p_j}}\right>, \\ \nonumber &=& {\gamma_c
e^{i\theta_{c}}} \left<{1 \over 1-\kappa_c -\kappa_{p_j}}\right>
\equiv {\overline g_c},\\ 
\eea 
\be \sigma^2_{\overline g_I} =
{\sigma^2_{\overline g_S} \over 2} + {\sigma^2_{\overline g_{p_j}}
\over 2}, 
\ee 
where 
\bea
\sigma^2_{g_I}\,\approx\,{\sigma^2_{p(\tau_S)}\over 2 N_{bg} } + {
\sigma^2_{\overline {g}_{p_j}} \over 2 N_{bg} }\,\approx\,
{\sigma^2_{p(\tau_S)}\over 2 N_{bg}} 
\eea 
${\sigma^2_{p(\tau_S)}}$
being the width of the intrinsic ellipticity distribution of the
sources, $N_{bg}$ the number of background galaxies averaged over and
$\sigma^2_{\overline {g}_{p_j}}$ the dispersion due to perturber
effects which should be smaller than the width of the intrinsic
ellipticity distribution. A more apt choice of coordinate system, the
polar $(u,v)$ provides the optimal measure. On averaging out the
smooth component, we have in polar coordinates: 
\bea 
\nonumber \left<{\overline
g_I}\right>_{uv} &=& \left<{\overline g_S}\right> + \left<{{\overline
\gamma_c} \over 1-\kappa_c - \kappa_{p_j}}\right> +
\left<{\gamma_{p_j}\over 1-\kappa_c -\kappa_{p_j}}\right>, \\
\nonumber &=& {\gamma_{p_j}}\left<{ 1 \over {1-\kappa_c
-\kappa_{p_j}}}\right>\equiv g_{p_ j},\\ 
\eea 
\bea
\left(\sigma^2_{\overline{g_I}}\right)_{uv} = {{\sigma^2_{\overline
g_S}} \over 2} + {{\sigma^2_{\overline g_c}} \over 2}, 
\eea 
where 
\bea
\sigma^2_{g_I}\,\approx\,{\sigma^2_{p(\tau_S)}\over 2 N_{bg} } +
{\sigma^2_{\overline{g}_{c}} \over 2 N_{bg} }.  
\eea 
From these
equations, we clearly see the two effects of the contribution of the
smooth cluster component: it boosts the shear induced by the perturber
due to the ($\kappa_c+\kappa_{p_j}$) term in the denominator, which
becomes non-negligible in the cluster center, and it simultaneously
dilutes the regular galaxy-galaxy lensing signal due to the
${\sigma^2_{\overline g_c} / 2}$ term in the dispersion of the
polarization measure. However, one can in principle optimize the noise
in the polarization by `subtracting' the measured cluster signal 
${\overline g_c}$ using
a fitted parametric model for the cluster and averaging in polar coordinates: 
\be
\left<{\overline g_I}-{\overline g_c}\right>_{uv} =
\left<{\gamma_{p_j}\over 1-\kappa_c -\kappa_{p_j}}\right>, 
\ee 
which
gives the same mean value as above but with a reduced dispersion: 
\be
\left(\sigma^2_{\overline g_I-\overline g_c}\right)_{uv} =
{\sigma^2_{\overline g_S} \over 2}, 
\ee 
where 
\bea
\sigma_{g_S}^2\,\approx\, {\sigma^2_{p(\tau_S)}\over 2 N_{bg}}.  
\eea
This subtraction of the larger-scale component reduces the noise in
the polarization measure, by about a factor of two; when
$\sigma^2_{\overline g_S}\sim \sigma^2_{\overline g_c}$, which is the
case in cluster cores. This differenced averaging prescription for
extracting the distortions induced by the possible presence of dark
halos around cluster galaxies is very feasible with {\it HST} quality data 
as we have shown in earlier work (Natarajan et al.\ 1998, 2002a).

Note here that it is the presence of the underlying large-scale smooth
mass distribution (with a high value of $\kappa_c$) that enables the
extraction of the weak signal riding on it. It is instructive to keep
in mind that in the regimes of interest discussed here the distortion
induced by the cluster-scale smooth component for a PIEMD model in the
inner-most (with a velocity dispersion of $1000\,{\rm kms}^{-1}$ and at
$R/r_t \leq 0.1$) regions is typically of the order of 20 - 40\% or so
in background galaxy shapes, and the perturbers produce distortions
(smaller scale PIEMDs with a velocity dispersion of $220\,{\rm kms}^{-1}$)
of the order of 5 - 10\%, significantly more than in the case of
weak-lensing by large scale structure or cosmic shear wherein the
distortions are of the order of 1\% percent.

\subsection{Modeling the cluster}

Each of the clusters studied in this paper preferentially probes the
high mass end of the cluster mass function and has a surface
mass density in the inner regions which is higher than the critical
value, therefore producing a number of multiple images of background
sources. By definition, the critical surface mass density for strong lensing 
is given by:
\begin{eqnarray}
\Sigma_{\rm crit} = {\frac{c^2}{4 \pi G}} \frac{D_s}{D_d D_{ds}}
\end{eqnarray}
where $D_s$ the angular diameter distance between the observer and the
source, $D_d$ the angular diameter distance between the observer and
the deflecting lens and $D_{ds}$ the angular diameter distance between
the deflector and the source. When the surface mass density in the
cluster is in excess of this critical value, strong lensing phenomena
with high magnification are observed. 

In general two types of lensing effects are produced -- strong:
multiple images and highly distorted arcs; and weak: small distortion
in background image shapes determined by the criticality of the
region. Viewed through the central, dense core region of the mass
distribution, where $\kappa > 1$ strongly lensed features are
observed. Note that the integrated lensing signal detected is due to
all the mass distributed along the line of sight in a cylinder
projected onto the lens plane. In this and all other cluster lensing
work, the assumption is made that individual clusters dominate the
lensing signal as the probability of encountering two massive rich
clusters along the same line-of-sight is extremely small due to the
fact that these are very rare objects in hierarchical structure
formation models.

With our current sensitivity limits, galaxy-galaxy lensing
within the cluster is primarily a tool to determine the total enclosed
mass within an aperture. We lack sufficient sensitivity to constrain
the detailed mass profile for individual cluster galaxies. With higher
resolution data in the near future we will be able to obtain
constraints on the slopes of mass profiles in sub-halos. In this
paper, we therefore concentrate on pseudo-isothermal elliptical
components (PIEMD models, derived by Kassiola \& Kovner 1993)
appropriately scaled for both the main cluster and the
substructure. We find that the results obtained for the
characteristics of the sub-halos (or perturbers) is largely
independent of the form of the mass distribution used to model the
smooth, large-scale component. A detailed comparison of the best-fit
profiles from lensing directly with those obtained in high resolution
cosmological N-body simulations is outside the scope of this paper and
will be presented elsewhere.

To quantify the lensing distortion induced by the global potential,
both the smooth and individual galaxy-scale halos are modeled
self-similarly using a surface density profile, $\Sigma(R)$ which is a
linear superposition of two PIEMD distributions,
\begin{eqnarray}
\Sigma(R)\,=\,{\Sigma_0 r_0  \over {1 - r_0/r_t}}
({1 \over \sqrt{r_0^2+R^2}}\,-\,{1 \over \sqrt{r_t^2+R^2}}),
\end{eqnarray}
with a model core-radius $r_0$ and a truncation radius $r_t\,\gg\,
r_0$. Correlating the above mass profile with a typical de
Vaucleours light profile (the observed profile for bright early type
galaxies) provides a simple relation between the truncation radius and
the effective radius $R_{\rm e}$, $r_t\sim (4/3) R_{\rm e}$. These
parameters $(r_t,r_0)$ are tuned for both the smooth component and the
perturbers to obtain mass distributions on the relevant scales. The
coordinate $R$ is a function of $x$, $y$ and the ellipticity, 
\bea
R^2\,=\,({x^2 \over (1+\epsilon)^2}\,+\,{y^2 \over (1-\epsilon)^2})\,;
\ \ \epsilon= {a-b \over a+b}, 
\eea 
The mass enclosed within radius $R$ for the $\epsilon = 0$ model is given by: 
\be 
M(R)={2\pi\Sigma_0
r_0 \over {1-{{r_0} \over {r_t}}}}
[\,\sqrt{r_0^2+R^2}\,-\,\sqrt{r_t^2+R^2}\,+\,(r_t-r_0)\,].  
\ee 
One of the attractive features of this
model is that the total mass ${M_{\infty}}$, is finite ${M_{\infty}}
\propto \,{\Sigma_0} {r_0} {r_t}$. Besides, 
analytic expressions can be obtained for the all the quantities of
interest, $\kappa$, $\gamma$ and $g$, \bea
\kappa(R)\,=\,{\kappa_0}\,{{r_0} \over {(1 - {r_0/r_t})}}\, ({1 \over
{\sqrt{({r_0^2}+{R^2})}}}\,-\,{1 \over
{\sqrt{({r_t^2}+{R^2})}}})\,\,\,, \eea \bea
2\kappa_0\,=\,\Sigma_0\,{4\pi G \over c^2}\,{D_{\rm ls}D_{\rm ol}
\over D_{\rm os}}, \eea where $D_{\rm ls}$, $D_{\rm os}$ and $D_{\rm
ol}$ are respectively the lens-source, observer-source and
observer-lens angular diameter distances which do depend on the choice
of cosmological parameters. To obtain $g(R)$, knowing the
magnification $\kappa(R)$, we solve Laplace's equation for the
projected potential $\phi_{\rm 2D}$, evaluate the components of the
amplification matrix and then proceed to solve directly for
$\gamma(R)$, and then $g(R)$. This yields for the projected potential,
\bea \phi_{2D}\,&=&\, \nonumber
2{\kappa_0}[\sqrt{r_0^2+R^2}\,-\,\sqrt{r_t^2+R^2}\,+ (r_0-r_t) \ln R\,
\\ \nonumber \,\, &-& \,r_0\ln\,[r_0^2+r_0\sqrt{r_0^2+R^2}]\,+
\,r_t\ln\,[r_t^2+r_t\sqrt{r_t^2+R^2}] ].\\ 
\eea 
And,
\bea
\gamma(R)\,&=&\,\nonumber \kappa_0[\,-{1 \over \sqrt{R^2 + r_0^2}}\, +\,{2 \over
R^2}(\sqrt{R^2 + r_0^2}-r_0)\,\\ \nonumber &+&\,{1 \over {\sqrt{R^2 +
r_t^2}}}\,-\, {2 \over R^2}(\sqrt{R^2 + r_t^2} - r_t)\,].\\ 
\eea
(then following equation 4, we can compute g(R)).
Scaling this relation by $r_t$ gives for $r_0<R<r_t$: 
\be
\gamma(R/r_t)\propto {\Sigma_0 \over \eta-1} {{r_t} \over
R}\,\sim\,{\sigma^2 \over R}, 
\ee 
where $\sigma$ is the velocity dispersion and for $r_0<r_t<R$: 
\be
\gamma(R/r_t)\propto {\Sigma_0\over\eta} {{r_t}^2 \over
R^2}\,\sim\,{{M_{\rm tot}} \over {R^2}}, 
\ee 
where ${M_{\rm tot}}$ is the total mass. In the limit that
$R\,\gg\,r_t$, we have, 
\bea 
\gamma(R)\,=\,{{3 \kappa_{0}} \over {2
{R^3}}}\,[{r_{0}^2}\,-\,{r_{t}^2}]\,+\,{{2 {\kappa_0}} \over {R^2}}
[{{r_t}\,-\,{r_0}}], 
\eea 
and as ${R\,\to\,\infty}$, $\gamma(R)\,\to\,0$, $g(R)\,\to\,0$ and
$\tau(R)\,\to\,0$ as expected.

Additionally, in order to relate the light distribution to key
parameters of the mass model above, we adopt a set of physically
motivated scaling laws for the cluster galaxies (Brainerd et al.\
1996):
\begin{eqnarray}
{\sigma_0}\,=\,{\sigma_{0*}}({L \over L^*})^{1 \over 4};\,\,
{r_0}\,=\,{r_{0*}}{({L \over L^*}) ^{1 \over 2}};\,\,
{r_t}\,=\,{r_{t*}}{({L \over L^*})^{\alpha}}.
\end{eqnarray}
These in turn imply the following scaling for the $r_t/r_0$ ratio $\eta$:
\bea
{\eta}\,=\,{r_t\over r_0}={{r_{t*}} \over {r_{0*}}} 
({L \over L^*})^{\alpha-1/2}.
\eea
The total mass $M_{\rm ap}$ enclosed within an aperture $r_{t*}$ and
the total mass-to-light ratio $M/L$ 
then scale with the luminosity as follows:
\begin{eqnarray}
M_{\rm ap}\,\propto\,{\sigma_{0*}^2}{r_{t*}}\,({L \over L^*})^{{1 \over
2}+\alpha},\,\,{M/L}\,\propto\,
{\sigma_{0*}^2}\,{r_{t*}}\left( {L \over L^*} \right )^{\alpha-1/2},
\end{eqnarray}
where $\alpha$ tunes the size of the galaxy halo and for $\alpha$ =
0.5 the assumed galaxy model has constant $M/L$ with luminosity (but
not as a function of radius) for each galaxy; if $\alpha>$ 0.5 ($\alpha<$
0.5) then brighter galaxies have a larger (smaller) halos than the
fainter ones. These scaling laws were empirically motivated by the
Faber-Jackson relation for early-type galaxies (Brainerd, Blandford \&
Smail 1996). We assume these scaling relations and recognize that this
could ultimately be a limitation but the evidence at hand supports the
fact that mass traces light efficiently both on cluster scales (Kneib et
al. 2003) and on galaxy scales (McKay et al. 2001; Wilson et al. 2001)
We also explore the dependence of the retrieved characteristic halo parameters on the
choice of the scaling index $\alpha$ for the most tightly constrained 
lens model in our sample, that of A\,2218..

\subsubsection{The intrinsic shape distribution of background galaxies}

As in all lensing work, it is assumed here as well, that the intrinsic or
undistorted distribution of shapes of galaxies is known. This
distribution is obtained from shape measurements taken from deep
images of blank field surveys. Previous analysis of deep survey
data such as the MDS fields (Griffiths et al.\ 1994) showed that the 
ellipticity distribution of sources is a strong function of the sizes 
of individual galaxies as well as their magnitude (Kneib et al.\
1996). For the purposes of our modeling, the intrinsic ellipticities
for background galaxies are assigned in concordance with an
ellipticity distribution $p_(\tau_S)$ where the shape parameter 
$\tau$ is defined as $\tau = (a^2-b^2)/(2ab)$ derived from the observed
ellipticities of the CFHT12k data (see Limousin et al.\ 2004 for details): 
\be
p(\tau_S)\,=\,\tau_S\,\,\exp(-({\tau_S \over
\delta})^{\nu});\,\,\,\nu\,=\,1.15,\,\,\delta\,=\,0.25.  
\ee
Note that this distribution includes accurately measured shapes of
galaxies of all morphological types. In the likelihood analysis
this distribution $p(\tau_S)$ is the assumed prior, which is used to
compare with the observed shapes once the effects of the assumed 
mass model are removed from the background images. We note here that 
the exact shape of the ellipticity distribution, i.e. the functional 
form and the value of $\delta$ and $\nu$ do not change the results, 
but alter the confidence levels we obtain. The width of the intrinsic
ellipticity distribution, on the other hand is the fundamental
limiting factor in the accuracy of all lensing measurements.

\subsubsection{The redshift distribution of background galaxies}

While the shapes of lensed background galaxies can be measured
directly and reliably by extracting the second moment of the light
distribution, in general, the precise redshift for each weakly object
is in fact unknown and therefore needs to be assumed. Using
multi-waveband data from surveys such as COMBO-17 (Wolf et al.\ 2004)
photometric redshift estimates can be obtained for every background
object. Typically the redshift distribution of background galaxies is modeled
as a function of observed magnitude $P(z,m)$. We have used data from
the high-redshift survey VIMOS VLT Deep Survey (Le Fevre et al.
2004) as well as recent CFHT12k R-band data to define the number counts of
galaxies, and the HDF prescription for the mean redshift per magnitude
bin, and find that the simple parameterization of the redshift
distribution used by Brainerd, Blandford \& Smail (1996) still
provides a good description to the data.

For the normalized redshift distribution at a given
magnitude $m$ (in the given band) we therefore have:
\bea 
N(z)|_{m}\,=\,{{\beta\,({{z^2} \over {z_0^2}})\, \exp(-({z \over
{z_0}})^{\beta})} \over {\Gamma({3 \over \beta})\,{{z_0}}}}; 
\eea
where $\beta\,=\,$1.5 and 
\bea z_0\,=\,0.7\,[\,{z_{\rm
median}}\,+\,{{d{z_{\rm median}}} \over
{dm_R}}{(m_R\,-\,{m_{R0}})}\,], 
\eea 
${z_{\rm median}}$ being the
median redshift, $dz_{\rm median}/ dm_R$ the change in median redshift
with say the $R$-band magnitude, $m_R$.

However, we note here in agreement with another recent study of
galaxy-galaxy lensing in the field by Kleinheinrich et al. (2004),
that the final results on the aperture mass are also sensitive
primarily only to the choice of the median redshift of the
distribution rather than the individual assigned values.

\subsection{The maximum-likelihood method}

Parameters that characterize both the global component and the
perturbers are optimized, using the observed strong lensing features -
positions, magnitudes, geometry of multiple images and measured
spectroscopic redshifts, when known, along with the smoothed shear
field as constraints. Note that from the above parameterization
presented in the previous section, it is clear that we can optimize
and extract values for $(\sigma_{0*}, r_{t*})$ for a typical $L^*$
cluster galaxy.

A maximum-likelihood estimator is used to obtain significance bounds
on fiducial parameters that characterize a typical $L^*$ sub-halo in
the cluster. We have extended the prescription proposed by Schneider
\& Rix (1996) for galaxy-galaxy lensing in the field to the case of
lensing by galaxy sub-halos in the cluster (Natarajan \& Kneib 1997,
Natarajan et al 1998). The likelihood function of the estimated
probability distribution of the source ellipticities is maximized for
a set of model parameters, given a functional form of the intrinsic
ellipticity distribution measured for faint galaxies.  For each
`faint' galaxy $j$, with measured shape $\tau_{\rm obs}$, the
intrinsic shape $\tau_{S_j}$ is estimated in the weak regime by
subtracting the lensing distortion induced by the smooth cluster model
and the galaxy sub-halos,
\begin{eqnarray}
\tau_{S_j} \,=\,\tau_{\rm obs_j}\,-{\Sigma_i^{N_c}}\,
{\gamma_{p_i}}\,-\, \gamma_{c}, 
\end{eqnarray}
where $\Sigma_{i}^{N_{c}}\,{\gamma_{p_i}}$ is the sum of the shear
contribution at a given position $j$ from $N_{c}$ perturbers. This
entire inversion procedure is performed for each cluster within the lens tool
utilities developed originally by Kneib (1993), which accurately takes 
into account the non-linearities
arising in the strong regime. Using a well-determined `strong lensing'
model for the inner-regions of the clusters derived from the positions,
shapes and magnitudes of the highly distorted
multiply-imaged objects 
along with the shear field determined from the shapes of the weakly distorted
background galaxies and assuming a known functional form for $p(\tau_{S})$
the probability distribution for the intrinsic shape distribution of
galaxies in the field, the likelihood for a guessed model is given by,
\begin{eqnarray}
 {\cal L}({{\sigma_{0*}}},{r_{t*}}) = 
\Pi_j^{N_{gal}} p(\tau_{S_j}),
\end{eqnarray}
where the marginalization is done over $(\sigma_{0*},r_{t*})$.
We compute ${\cal L}$ assigning the median redshift corresponding to the
observed source magnitude for each arclet. The best fitting model 
parameters are then obtained by maximizing the log-likelihood 
function $l$ with respect to the parameters ${\sigma_{0*}}$ and ${r_{t*}}$. 
Note that the parameters that characterize the smooth component are
also simultaneously optimized. The likelihood can also be marginalized over
a complementary pair of parameters, i.e.\ $\alpha$ the luminosity scaling
index and the aperture mass $M_{\rm ap}$ directly. In this work,
we explore both choices. 

\subsection{The mass function of substructure in the inner regions}

Mapping of the substructure mass function is ultimately of interest
since it intimately connects to the galaxy formation process.
Comparison of the dark halo mass function with the observed luminosity
function of galaxies provides valuable insights into the process of
assembly of galaxies. Substructures in the dark matter are defined to
be lower mass clumps that are dynamically distinct, and bound objects
that reside inside a virialized dark matter halo. The existence of
substructure is a generic prediction of hierarchical structure
formation in CDM models. The assembly of collapsed mass in these
models proceeds via the gravitational amplification of initial density
fluctuations resulting in dark halos that are not smooth and
structureless but rather clumpy with significant amounts of
substructure. For instance, within a radius of $400\,h^{-2}\,{\rm
kpc}$, from the Milky Way, cosmological models of structure formation
predict $\sim$ 50 dark matter satellites with circular velocities in
excess of $50\,{\rm kms^{-1}}$ and mass greater than
$3\,\times\,10^8\,M_{\odot}$.  This number is significantly higher
than the dozen or so satellites actually observed around our
Galaxy. If we extend the analysis to the Local Group the problem gets
worse, as $\sim$ 300 satellites are predicted inside a $1.5\,{\rm
Mpc}$ radius whereas only about 40 are detected. However, the observed
VDF (velocity distribution function) and the predicted VDF do match up
at $v_{\rm circ} = 50\,{\rm kms^{-1}}$, indicating that the abundance
of satellites is a problem on small scales (Klypin et al.\ 2002). On
small scales, $M\,<\,10^{8}\,M_{\odot}$ or so, there is a marked
discrepancy between the theoretical/N-body simulation prediction for
the amount of substructure that is observed in the Universe. A surfeit
of sub-halos (satellites) are predicted for a galaxy like the Milky
Way, when only a handful are detected. There are several possible
explanations for this discrepancy including (i) possible
identification of some satellites with the detected High Velocity
Clouds (HVCs) and (ii) physical processes inhibiting star formation
preferentially in low mass halos implying the existence of large
numbers of dark satellites. If the vast number of sub-halos are dark,
lensing might be the best way to detect them.

Recently Lee (2004) has attempted an analytic calculation of the
sub-halo mass function inside clusters for CDM models and finds that
$n(M)\,\propto\,{M^{-0.8}}$ over the mass range
$10^{11}$--$10^{12.5}\,M_{\odot}$. Lee (2004) takes into account the
complex dynamical history of galaxies in the cluster using one
parameter to model the effect of global tidal truncation. Therefore,
adopting the simple tidal-limit approximation to estimate analytically
the global and local mass distribution of dark matter halos that
undergo tidal mass-loss, Lee (2004) finds that the resulting mass
functions are in excellent agreement with what has been found in
recent N-body simulations. We have also demonstrated that the spatial
extent inferred from galaxy-galaxy lensing are consistent with the
tidal stripping hypothesis (see Natarajan et al.\ 2002a).  Similar to
our results here, Lee also finds that only about 10\% of the sub-halo
mass is in this mass range. Numerical studies by De Lucia et al.\
(2004) find that the substructure mass function depends only weakly on
the properties of the parent halo mass, and is well described by a
power law. The mass fraction in substructure also appears to be 
relatively insensitive to the tilt and overall normalization of the 
primordial power spectrum (Zentner \& Bullock 2003).

\section{Lensing Analysis of Clusters}  

We briefly discuss the properties of the five lensing clusters that
are studied here. The clusters in order of increasing redshift are:
A\,2218, $z = 0.18$; A\,2390, $z = 0.23$; Cl\,2244$-$02, $z = 0.33$;
Cl\,0024+16, $z = 0.39$; Cl\,0054$-$27, $z = 0.57$.  All clusters have
multiply imaged background sources (some with several sets of
multiply-imaged sources) with measured spectroscopic redshifts that
are used to calibrate the overall mass model.  In addition to these
five clusters we will also include the results from our previous
analysis of the rich cluster AC\,114 at $z=0.31$ described in
Natarajan et al.\ (1998).  The {\it HST} WFPC2 imaging of this cluster
was analyzed and modeled in an identical manner to that used here and
hence allowing those results to be included in our discussion.

The X-ray and lensing properties of three of the clusters analyzed
here (A\,2218, Cl\,0024+16 and Cl\,0054$-$27) were discussed by Smail
et al.\ (1997b) based on the data available at that time.  We
summarize the information from more recent observations of these
clusters, as well as the two remaining systems (A\,2390 and
Cl\,2244$-$02) below.  The clusters range over an order of magnitude
in terms of their X-ray luminosity ($L_{\rm X} \sim 10^{43-44} h^{-2}$
erg s$^{-1}$) and roughly an order of magnitude in terms of their
$V$-band luminosities ($L_{\rm V} \sim $0.25--1$ \times10^{12}\,h^{-2}
L_{\odot}$).

While these {\it HST} cluster-lenses span a large range in mass,
richness, and X-ray luminosity, fortunately, they form a subset of
clusters with morphologically well-studied galaxy populations (Couch
et al.\ 1998; Smail et al.\ 1997a). For four of the clusters studied 
the morphological classification for the cluster members and cluster
membership was obtained from the following sources: for AC\,114 from 
Couch et al.\ (1998), for Cl\,0024+16 and Cl\,0054$-$27 from 
(Smail et al.\ 1997a). We used only color-selection to determine
cluster membership and classification for A\,2390 and Cl\,2244-02.

\subsection{The HST cluster lens sample}

\subsubsection{A\,2218}

A\,2218 is one of the best-studied cluster lenses, with over 7
multiply-imaged background sources identified by {\it HST} observations 
(Kneib et al.\ 1996, 2004a, 2004b, Ellis et al 2001).  The core of the
cluster is dominated by a luminous cD galaxy and the galaxy population
within the central 1\,h$^{-1}$\,Mpc is made up predominantly of
morphologically-classified early-type galaxies (Couch et al.\ 1998;
Zeigler et al.\ 2001).  {\it Chandra} observations of A\,2218 yield a
mean cluster temperature of $kT = 6.9 \pm 0.5\,{\rm keV}$ and a rest-frame
luminosity in the 2--10\,keV energy band of $6.2\,\times\,10^{44}\,
{\rm ergs\,s^{-1}}$ (Mahacek et al.\ 2002). The high-resolution {\it
Chandra} data of the inner $2'$ of the cluster show that the X-ray
brightness centroid is apparently displaced in projection from the cD
galaxy. Asymmetric temperature variations are also detected along the
direction of the cluster mass elongation.  Although the X-ray and weak
lensing mass estimates are in good agreement for the outer parts
($r\,>\,200\,h^{-1}$\,kpc) of the cluster, in the inner region the
observed X-ray temperature distribution is inconsistent with the
assumption of the intra-cluster gas being in thermal hydrostatic
equilibrium, pointing to recent merger activity.

\subsubsection{A\,2390}

A\,2390, at $z = 0.23$, is extremely luminous and hot in the X-rays.
Recent {\it Chandra} measurements by Allen, Ettori \& Fabian (2001)
find an isothermal temperature distribution between $200\,h^{-1}\,{\rm
kpc}$ and $1\,{\rm Mpc}$ with $kT\,=\,11.5\,{\rm keV}$, with a decline in
the temperature within $200\,h^{-1}\,{\rm kpc}$.  This rich cluster
has a significant early-type galaxy population that is concentrated in
the inner regions (Fritz et al.\ 2003). The X-ray surface brightness
profile is smooth and the optical data also suggest that the cluster
is in dynamical equilibrium. The projected mass profile obtained from lensing,
optical measurements of the galaxy velocity dispersions and the X-ray
data from Chandra are in good agreement (Allen, Ettori \& Fabian
2001).

\subsubsection{Cl\,2244$-$02}

Cl\,2244$-$02 is a very compact cluster at $z=0.33$ which produces a
near complete Einstein ring image of a background galaxy at $z=2.237$
(Lynds \& Petrosian 1989; Hammer et al.\ 1989; Mellier et al.\ 1991),
as well as a near-infrared selected giant arc (Smail et al.\ 1993).
This remarkable lensing configuration confirms the massive mass
concentration in the central regions of this cluster.  The {\it HST}
WFPC2 image reveals a very concentrated distribution of early-type
galaxies in the inner $30\,{\rm arcseconds}$, surrounded by the
Einstein ring, however, there are relatively few obvious cluster
galaxies outside this region.  Moreover, the {\it ASCA} X-ray
observations of the cluster by Ota et al.\ (1998) gives $kT = 6.5 \pm
1.3\,{\rm keV}$ and a rest-frame luminosity of just $1.3\,\times\,10^{44}\,
{\rm erg\,s^{-1}}$ in the 2--10\,keV band.  The relatively low X-ray
luminosity suggests that the cluster has a low mass, although the
X-ray temperature indicates a more massive system.  This is supported
by the lensing model we have constructed for Cl\,2244$-$02.

\subsubsection{Cl\,0024+16}

The rich cluster Cl\,0024+16 at $z = 0.39$ has a measured X-ray
temperature of just $kT \sim 4.5\,{\rm keV}$ (Ota et al.\ 2004). Ota
et al.\ (2004) find that the surface brightness profile is represented
by the sum of extended emission centered at the central bright
elliptical galaxy with a small core of $\sim\,50\,{\rm kpc}$ and more
extended emission with a core radius of $\sim\,210\,{\rm kpc}$.
However, this was one of the cases where the mass determinations from
three independent techniques: lensing, using virial estimators and
from the X-ray data under the assumption of hydro-static equilibrium
for the gas were highly discrepant. Using spectroscopic information
for about 300 galaxies within a projected radius of $3\,{\rm
h^{-1}Mpc}$ Czoske et al.\ (2002) examined the three-dimensional
structure of this cluster and found that dynamically there were two
distinct components separated by $\sim\,3000\,{\rm kms^{-1}}$ in
velocity space. They argue that this is suggestive of a high-speed
collision between these two sub-clusters. Such an interpretation would
explain the origin of the disagreement between the various mass
estimates. Recently published work by Kneib et al.\ (2003) using a
panoramic sparsely sampled image from WFPC2 and STIS on {\it HST}
derive a best-fit mass model from the lensing data out to $5\, {\rm
h^{-1}Mpc}$ in which they identified a secondary mass clump with about
30\% of the overall cluster mass. Note that in this paper we construct
a high-resolution mass model only for the inner region.

\subsubsection{Cl\,0054$-$27}

The most distant cluster in our sample is Cl\,0054$-$27 at $z =
0.57$. This is an optically-selected cluster with a dominant central galaxy (Couch et
al.\ 1985). A multiply-imaged arc is visible in the {\it HST} images
of this cluster with a measured a redshift of 3.2 for this feature
(Leborgne, private communication). The X-ray luminosity of this
cluster measured to be $2.5\,h^2\,10^{43}\,{\rm erg\,s^{-1}}$ in the 0.3 --
3.5\,keV band, is lower than would be expected from the $L_X$ to the
measured shear strength correlation for massive lensing clusters (see
Fig.~2 of Smail et al. 1997 for the correlation between $L_X$ and
$<g>$ for a sample of HST cluster-lenses). Smail et al. (1997) argue
that this cluster is an example of a system that is elongated along
the line-of-sight, leading to the low $L_X$ for the measured surface
mass density. Obviously, in cases like this the mass estimate obtained
from X-ray data which assumes spherical symmetry is unlikely to
provide accurate results.

\subsection{Lensing Constraints}

There are two aspects to constructing a successful lens model for the
clusters analyzed here.  Firstly, we must identify multiply-imaged
background sources with reliable redshift measurements whose
properties can be used to constrain the total projected mass within
the lens models.  Secondly, we have to extend these models to larger
radii using the coherent distortion signal induced in the shapes of
faint, background galaxies by the foreground cluster potential well.

Both of these steps use the deep, high-resolution imaging provided for
the cluster cores by{\it HST}.  All five clusters analyzed here were
observed with WFPC2 on-board {\it HST} for Guest Observer programs
GO 5352 (A\,2390 and Cl\,2244$-$02), 5378 (Cl\,0054$-$27) 5453
(Cl\,0024+16) and 5701 (A\,2218).  The filters used for the
observations were F555W ($V_{555}$) and F814W ($I_{814}$) or F702W ($R_{702}$).
Observations of A\,2218 were taken in the $R_{702}$ filter and
all the other clusters studied here were observed in the $I_{814}$ filter. 
We have used the color information, when available, to test the
identification of multiply-imaged sources in these fields. However, in
the following analysis we use the reddest band available for a
particular cluster to catalog objects and measure their shapes.  The
total exposure times are then 10.5\,ks on both A\,2390 and
Cl\,2244$-$02, 16.8\,ks on Cl\,0054$-$27 and 13.2\,ks on Cl\,0024+16,
all in $I_{814}$, and 6.3\,ks in $R_{702}$ on A\,2218.  The individual
exposures were generally grouped in sets of four single-orbit
exposures each offset by 2.0 arcsec to allow for hot pixel
rejection. After standard pipeline reduction, the images were aligned
using integer pixel shifts and combined into final frames using the
{\sc iraf/stsdas} task {\sc crrej}.  We retain the WFPC2 color system
and hence use the zero points from Holtzman et al.\ (1995). The final
images cover the central 0.8--1.6 Mpc of the clusters (Fig.~1--5) to a
5-$\sigma$ point-source limiting magnitude of $I_{814} \sim 26.0$ or
$R_{702} \sim 26.5$--27.0 (Smail et al.\ 1997a).

Multiply-imaged background galaxies have been identified using {\it
HST} imaging and spectroscopically confirmed in A\,2218 by Kneib et
al.\ (1996, 2004b), in A\,2390 by Pello et al.\ (1999) and Frye \&
Broadhurst (1998), Cl\,2244$-$02 (Smail et al.\ 1995a; Mellier et al.\
1991), Cl\,0024+16 by Broadhurst et al.\ (2000) and Cl\,0054$-$27 by
LeBorgne (priv. communication).  These provide very strong constraints
and drive the fit in the likelihood plane as we discuss below.

The second step in our analysis requires a statistical measure of the
shear induced in the background field population. To achieve this we
catalog faint objects in these frames and measure their shapes using
the {\sc Sextractor} image analysis package (Bertin \& Arnouts 1995).
Following Smail et al.\ (1997a, 1997b) we adopt a detection isophote
equivalent to $\sim 1.3 \sigma$ above the sky, where $\sigma$ is the
standard deviation the sky noise, e.g.\ $\mu_{814}= 25.0$ mag
arcsec$^{-2}$ or $\mu_{702} = 25.0$ mag arcsec$^{-2}$ for A\,2218, and
a minimum area after convolution with a 0.3 arcsec diameter top-hat
filter of 0.12 arcsec$^{2}$.  Analysis of our exposures provides
catalogs of $\sim 800$ objects for each cluster across the 3 WFC
chips.  We discard the smaller, lower sensitivity, PC fields as well
as a narrow border around each WFC frame in the following analysis.
The imaging data (and catalogs) used here are identical to those
previously analysed by Kneib et al.\ (1996) for A\,2218, and Smail et
al.\ (1997a, 1997b) for A\,2218, Cl\,0024+16 and Cl0054$-$27.
In order to account for the error in shapes produced due to PSF, we
adopt a method similar to that of Smith et al.\ (2004). 

To constrain the weak lensing aspect of the models of the various
clusters we must construct well-defined samples of background galaxies
for which image parameters can be measured with adequate
signal-to-noise. For simplicity in modeling we have adopted uniform
magnitude limits across the sample. The faint magnitude limit is
determined by the depth at which reliable images shapes can be
measured in our shortest exposures. This is $R_{702} = 26.0$, as set
by the A\,2218 exposure.  The bright limit is set by our desire to
reduce cluster galaxy contamination in the field samples for the most
distant clusters and corresponds to a bright limit of $I_{814} =
22.5$.  When converting between the $R_{702}$ and $I_{814}$ limited
samples, we have assumed a typical color for the faint field
population at these depths of $(R_{702}\! -\!  I_{814}) \sim 0.5$
(Smail et al.\ 1995b).  Hence our background galaxy cut is defined
simply as $R_{702}=23$--26 or $I_{814}=22.5$--25.5.  Cluster galaxies
are chosen with an additional luminosity cut-off.

Applying these limits yields a typical surface density of $\sim 95$
field galaxies per arcmin$^2$, in good agreement with that measured in
genuine `blank' fields ($\sim 95\pm10$ arcmin$^{-2}$) after correcting
for differences in the photometric systems (Smail et al.\ 1995b). We thus
estimate that any residual contamination in our catalogs from faint
cluster members must be less than $\sim 5$--10\%. The final sample
size in a typical cluster, after applying both the magnitude and the
area cuts (see Table~3), is $\sim 350 - 400$ galaxies.

To determine the contribution to the observed shear from systematic
effects in the {\it HST} optics, detectors, or the reduction method, we have
also modeled  the PSF anisotropy using the methods adopted by Smith et
al. (2004) and corrected the background image
shapes accordingly.

\subsection{Detailed mass models}

A composite mass model is constructed for the clusters starting with
the super-posed PIEMDs. The strong lensing data, i.e. the geometry,
positions, relative brightness, redshifts and parities of the multiple
images are used to obtain the mass enclosed within the Einstein radius which
is used as an initial constraint for the integrated mass in the inner
regions. The contribution to the shear and magnification from all
potentials (large scale and small-scale perturbers) is calculated at
the location of every observed background source galaxy and the
inversion of the lensing equation is performed. The observed shape and
magnification of each and every distorted background galaxy is
compared to that computed from the model and the sub-halo mass
distribution is modified iteratively till the best match between the
observations and the model are found simultaneously for all background
sources.

The basic steps of our analysis involves the lens inversion, modeling and
optimization, which are done using the {\sc lenstool} software utilities
(Kneib 1993). These utilities are used to perform the ray tracing from
the image plane to the source plane with a specified intervening
lens. This is achieved by solving the lens equation 
iteratively, taking into account the observed strong lensing features,
positions, geometry and magnitudes of the multiple images. 
In some cases, we also include a constraint on the location of the critical 
line (between 2 mirror multiple images) to fasten the optimization.
In Figs.~1--5, we show the iso-mass contours overlaid on their
respective {\it HST} WFPC2 images. All the cluster galaxies included
in the analysis have ellipses around them, and over-plotted are the
critical curves (in yellow) for three different source redshifts ($z_s = 1, 2,
3$), the multiple images (in cyan) and the smoothed background shear
field (in magenta) for the best-fit model. Additionally, we fix the 
core radius of an $L^*$ sub-halo to be $0.1\,{\rm kpc}$, as by construction
our analysis cannot constrain this quantity. In addition to the
likelihood contours, the reduced $\chi^2$ for the best-fit model is
also robust. We describe pertinent features of each cluster and their
respective mass models below.

\noindent{\bf A2218}

Our best fit mass model for the cluster is bimodal, composed of two
large scale clumps around the cD and the second brightest cluster
galaxy (Fig.~1 and Table~1).  This model is an updated version of that
constructed by Kneib et al.\ (1996). It includes 40 additional
small-scale clumps that we associate with luminous early-type galaxies
in the cluster core. Only about 10\% of the total cluster mass is in
the substructure i.e.\ associated with galaxy scale halos. The aperture
mass, integrated over the truncation radius $r_{ap}\,=\,40$ kpc,
yields a characteristic mass of $1.4 \times 10^{12}\,\msun$, with a
total mass-to light ratio in the V-band of $\sim\,5.8 \pm 1.5$ and a central
velocity dispersion of about $180\,{\rm km\,s}^{-1}$.

\begin{table*}[ht!] 
\begin{center}
\begin{tabular}{lccccccc}
\hline\hline\noalign{\smallskip}
${z}$& $x$ & $y$ & $\epsilon$ & ${\theta}$ & ${\sigma}$ & ${r_t}$ &$ {r_c}$\\

& & {\rm arcsec} & {\rm arcsec} & {\rm deg} & (km\,s$^{-1}$) & (\rm kpc) &
(\rm kpc)\\
\noalign{\smallskip}
\hline
 \noalign{\smallskip}
{\bf A\,2218}\\
${0.17}$ & ${0.3}$ & ${1.4}$ & ${0.3}$ & ${-13}$ & ${1070}$ & ${900}$ 
& ${75}$\\
${0.17}$ & ${-67.5}$ & ${3.0}$ & ${0.2}$ & ${20}$ & ${400}$ & ${600}$ 
& ${25}$\\
\noalign{\smallskip}
{\bf A\,2390}\\
${0.23}$ & ${0.0}$ & ${0.0}$ & ${0.1}$ & ${17}$ & ${1100}$ & ${900}$ 
& ${55}$\\
${0.23}$ & ${-0.04}$ & ${-2.70}$ & ${0.4}$ & ${16}$ & ${450}$ & ${60}$ 
& ${10}$\\
\noalign{\smallskip}
{\bf Cl\,2244$-$02}\\
${0.33}$ & ${0.0}$ & ${0.0}$ & ${0.17}$ & ${45}$ & ${600}$ & ${900}$ 
& ${30}$\\
${0.33}$ & ${17.32}$ & ${-10.2}$ & ${0.1}$ & ${90}$ & ${300}$ & ${600}$ 
& ${20}$\\
\noalign{\smallskip}
{\bf Cl\,0024+16}\\
${0.39}$ & ${0.3}$ & ${1.4}$ & ${0.3}$ & ${-13}$ & ${1000}$ & ${900}$ 
& ${30}$\\
${0.39}$ & ${1.63}$ & ${71.3}$ & ${0.3}$ & ${45}$ & ${200}$ & ${200}$ 
& ${20}$\\
\noalign{\smallskip}
{\bf Cl\,0054$-$27}\\
${0.57}$ & ${0.0}$ & ${-1.0}$ & ${0.2}$ & ${-22}$ & ${1100}$ & ${900}$ 
& ${30}$\\
\hline
\end{tabular}
\end{center}
\caption{Properties for the primary and 
(where relevant) secondary mass clumps in the
clusters. The characteristic parameters are initially constrained by 
the positions, shapes and luminosities of the multiple-imaged 
objects and are then iteratively varied to match the weak shear field
as well to obtain the optimal values (in the 
$\chi^2$ sense) in a likelihood scheme.}
\end{table*}

Additionally, we explore the relation between luminosity and mass (velocity
dispersion, in fact) by allowing the index $\alpha$
to float. This is done by using an alternate choice of parameters to
construct the likelihood function ${\cal L(\alpha, M_{\rm ap})}$.
The results of this analysis are  shown in Fig.~9. The retrieved value
of the aperture mass $M_{\rm ap}$ is not particularly sensitive to the
choice of $\alpha$.

\noindent{\bf A2390}

The cluster has an unusual feature -- a strongly lensed almost
`straight arc' (Pello et al.\ 1991) approximately 38 arcsec ($\sim$
170 kpc) away from the central galaxy, in addition to many other arcs
and arclets that have been utilized in our modeling. We find a
best-fit mass model with two large-scale components (see Table~1 for
their properties), that yield a projected mass within the radius
defined by the brightest arc of $\sim 1.8 \pm 0.2 \times
10^{14}\,M_\odot$.  Our best-fit composite lensing model for A\,2390
incorporates 40 perturbers associated with early-type cluster members
whose characteristic parameters are optimized in the
maximum-likelihood analysis.  We show the equi-potentials of this mass
model overlaid on the {\it HST} WFPC2 data in Fig.~2, showing those
cluster galaxies selected as perturbers and we also show the critical
curves for three different source redshifts ($z_s = 1, 2, 3$), the multiple
images used to constrain the model and and the smoothed background
shear field from the best-fit model. The integrated mass within the
$\sim$ 18 kpc tidal radius for a typical  $L^*$ cluster galaxy is
about $6.4 \times 10^{11}\,\msun$ giving a total mass-to-light ratio
in the V-band of about $4.2 \pm 1.3$. Once again 90\% of the total
mass of the cluster is consistent with being smoothly distributed.

\noindent{\bf Cl2244}

This best-fit lens model for this cluster has two components but both have 
fairly low velocity dispersions (see Table~1). This is the least
massive lensing cluster in the sample studied here. The X-ray mass
estimate from the {\it ASCA} data Ota et al.\ (1998) is in good
agreement with our best-fit lensing mass model. This is despite the fact that
the X-ray temperature of Cl\,2244$-$02 is at least a factor of two higher 
than that expected from the average luminosity-temperature relation.

The tidal truncation radius obtained for a typical $L^*$ cluster
galaxy in Cl\,2244 is the largest in the sample studied here and is
$55 \pm 12$ kpc. This is in consonance with the fact that the central
density in Cl\,2244 is the lowest. The total mass-to-light ratio in
the V-band for a fiducial $L^*$ is $3.2 \pm 1.2$. Approximately 20\%
of the total mass is in substructure within the mass range $10^{11} -
10^{12.5}\,\msun$.

\noindent{\bf Cl0024}

Our best fit mass model for the inner regions takes into account the small
scale dark halos associated with the early-type members in the core,
and requires a two component model for the sub-clusters (Table~1 and Fig.~4).  
Integrating the best-fit mass model shown in Fig.~4, we find that
(i) about 10\% of the total cluster mass is in galaxy-scale halos and
(ii) the total mass estimate is in good agreement with that obtained
by Kneib et al.\ (2003) where data from a much larger field of view was used.

Even on the large scales probed by Kneib et al.\ (2003) it was found
that mass and light traced each other rather well at large
radii. A typical $L^*$
cluster galaxy was found to have  a truncation radius of $45 \pm 5$ kpc, and a
central velocity dispersion of $125 \pm 7\,{\rm km\,s}^{-1}$.

\noindent{\bf Cl0054}

The lensing signal from Cl\,0054$-$27 is best fit by a single smooth
dark matter component and sub-halos associated with bright,
early-type members making it the only uni-modal cluster in the sample
studied here. The mass enclosed within $\sim$ 400 kpc is of the order
of $1.8 \pm 0.4 \times 10^{14}\,\odot$. The best-fit mass model is
plotted in Fig.~5, with all the cluster galaxies included in the model
shown explicitly.

The characteristic central velocity dispersion of a typical $L^*$
galaxy in this cluster is higher than in A\,2218, A\,2390 or
Cl\,0024+16, all of which are by contrast bimodal in the mass
distribution. In this cluster about 20\% of the total mass is
in substructure. However,  Cl\,0054$-$27 is the most distant cluster 
studied here and is likely to be still evolving and assembling
accounting for the high mass fraction in substructure.

\begin{table*}[!ht]
\begin{center}
\begin{tabular}{lccc}
\hline \\
{\rm Cluster} & ${z}$ & ${\rm N_{\rm cg}}$ & ${\rm N_{\rm bg}}$ \\
\hline \\
{\bf A\,2218}     & ${0.17}$ & ${40}$ & ${358}$ \\
{\bf A\,2390}     & ${0.23}$ & ${40}$ & ${378}$ \\
{\bf Cl\,2244-02} & ${0.33}$ & ${40}$ & ${398}$ \\
{\bf Cl\,0024+16} & ${0.39}$ & ${40}$ & ${344}$ \\
{\bf Cl\,0054-27} & ${0.57}$ & ${40}$ & ${426}$ \\
\hline \\
\end{tabular}
\end{center}
\caption{The number of background galaxies ${\rm N_{\rm bg}}$ and 
foreground lenses ${\rm N_{\rm cg}}$ for the clusters studied here. The number of background sources
was determined usign a magnitude cut and the lenses were picked
to be the 40 brightest, early-type galaxies from confirmed cluster 
members.}
\end{table*}

\begin{table*}
\begin{center}
\begin{tabular}{lcccccccc}
\hline\hline\noalign{\smallskip}
${\rm Cluster}$& & ${\sigma^\ast}$&${r_t^\ast}$&${M_{\rm ap}/L_v}$&$\rm {M^\ast}$&
$\sigma_{\rm clus}$ & ${\rho_{\rm clus}(r = 0)}$\\
& & (km\,s$^{-1}$) & (kpc) & 
(M$_\odot$/L$_\odot$) & (10$^{11}$M$_\odot$) & (km\,s$^{-1}$) &
(10$^6$ $\msun$ kpc$^{-3}$)\\
\noalign{\smallskip}
\hline
\noalign{\smallskip}
{A\,2218} & ${0.17}$ & ${180\pm10}$ & ${40\pm12}$ &
${5.8\pm1.5}$ & $\sim\,14 $ & ${1070\pm70}$  &  {3.95}\\

{A\,2390} & ${0.23}$ & ${200\pm15}$ & ${18\pm5}$ &
${4.2\pm1.3}$ & $\sim\,6.4 $  &${1100\pm80}$& {16.95}\\

{AC\,114}  & ${0.31}$ & ${192\pm35}$ & ${17\pm5}$ &
${6.2\pm1.4}$ & $\sim\,4.9 $ &${950\pm50}$& {9.12}\\ 

{Cl\,2244$-$02} & ${0.33}$ & ${110\pm7}$ & ${55\pm12}$ &
${3.2\pm1.2}$ & $\sim\,6.8 $ &${600\pm80}$  & {3.52}\\

{Cl\,0024+16} & ${0.39}$ & ${125\pm7}$ & ${45\pm5}$ &
${2.5\pm1.2}$ & $\sim\,6.3 $ &${1000\pm70}$ & {3.63}\\

{Cl\,0054$-$27} & ${0.57}$& ${230\pm18}$ & ${20\pm7}$ &
${5.2\pm1.4}$ & $\sim\,9.4 $ &${1100\pm100}$ & {15.84}\\
\noalign{\smallskip}
\hline
\end{tabular}
\end{center}
\end{table*}

As illustrated in Fig.~9, the derived mass is not a strong function of
$\alpha$ given the errors. However, from other lensing work notably by
McKay et al.\ (2001) and Wilson et al.\ (2001) it is clear that mass and
light seem to trace each other rather tightly on galactic scales
although in the inner-most regions of galaxies baryons appear to
dominate density profiles. Note here that the choice of $\alpha$
determines only the scaling of the outer radius of a fiducial sub-halo
with luminosity.  With the data used in this paper it is not possible
to distinguish between various values of $\alpha$ - some values are
clearly more physical than others.  Therefore, this
implies that we are sensitive to the integrated mass within an
aperture that is determined primarily by the anisotropy in the shear
field and not on the details of the how the sub-halo masses are
truncated. We also find that out to 500 kpc in all clusters only
10--20\% of the total mass is associated with galaxy halos, at this
radius (to which we are limited due to the size of the {\it HST} WFPC2
fields) most of the mass is in the large scale component. Needless
this fraction is likely to be a strong function of cluster-centric
radius. The dependence of the efficiency of tidal stripping with
distance from the cluster can be explored with wide-field HST data and
we are in the process of doing so for the cluster Cl\,0024+16
(Natarajan et al.\ 2004).

\section{Uncertainties}

\subsection{Systematic errors: robustness of the lens models}

The following tests were performed for each cluster, (i) choosing
random locations (instead of bright, early-type cluster member
locations) for the perturbers; (ii) scrambling the shapes of
background galaxies; (iii) choosing to associate the perturbers with
the 40 faintest (as opposed to the 40 brightest) galaxies; (iv)
randomly selecting known cluster galaxies as perturbers; (v) selecting
late-type galaxies. None of the above cases (i)-- (v) yields a
convergent likelihood map, in fact all that is seen in the resultant
2-dimensional likelihood surfaces is noise.

The robustness of our results has been amply tested, however there are
a couple of caveats that we ought to mention. As outlined above in
this galaxy-galaxy lensing technique we are sensitive to only a
restricted mass range in terms of secure detection of substructure.
This is due to the fact that we are quantifying a differential signal
above the average tangential shear induced by cluster, and we are
inherently limited on average by the number of distorted background
galaxies that lie within (1 - 2) tidal radii of cluster galaxies.
This trade-off between requiring sufficient number of lensed
background galaxies in the vicinity of the sub-halos and the optimal
locations for the sub-halos leads us to choose the brightest 40
early-type cluster galaxies for each lens. With deeper, wider and more
numerous images of clusters, expected in the future with a wide-field imager in
space, such as the SNAP mission (Aldering et al. 2003) this technique can be pushed much
further to probe down to lower masses in the mass function. It is
possible that the bulk of the mass in sub-halos are in lower mass
clumps (which in this analysis is essentially accounted for as part of
the smooth component) and are in fact anti-correlated with positions
of early-type galaxies.

Our results still hold true since we are filtering out only the most
massive clumps via this technique. Note that one of the null tests
performed above, associating galaxy halos with random positions in the
cluster (and not on the locations of bright, early-type galaxies)
resulted in pure noise. Even if we suppose that the bulk of the dark
matter is associated with say, dwarf/very low surface brightness
galaxies in clusters, then the spatial distribution of these galaxies
is required to be fine-tuned such that these effects do not show up in
the shear field in the inner regions implying that if at all they are likely to
be more significant repositories of mass perhaps in the outskirts of clusters.

Guided by the current theoretical understanding of the assembly of clusters,
dwarf galaxies are unlikely to survive in the high density core
regions of galaxy clusters studied here. When studies such as
presented here are applied on a larger scales to distances over a few
Mpc's from the cluster center (analysis of the mosaic-ed HST images as in
Natarajan et al.\ 2004) we can explore further the relation
between mass and light, and the variation of tidal truncation radii with
distance from the cluster center.

\subsection{Random errors}

The principal sources of error in the above analysis are (i) shot
noise -- we are inherently limited by the finite number of sources
sampled within a few tidal radii of each cluster galaxy; (ii) the
spread in the intrinsic ellipticity distribution of the source
population; (iii) observational errors arising from uncertainties in
the measurement of ellipticities from the images for the faintest
objects and (iv) contamination by foreground galaxies mistaken as
background. As mentioned in Section 2.2.2, the partitioning of mass
into sub-halos and the smooth component as done here is largely
independent of the $N(z)$ of background galaxies.

In terms of the total contribution to the error budget,
performing simulations we find that the shot noise is the most
significant source of error $\sim 50\%$; followed by the width of the
intrinsic ellipticity distribution which contributes $\sim 20\%$, and
the other three sources together contribute $\sim 30\%$. This
elucidates the future strategy for such analyses - going significantly
deeper and wider in terms of the field of view is likely to provide
considerable gains. Mosaic-ed ACS images are the ideal data sets for
this galaxy-galaxy lensing analyses, and such work is currently in
progress.

\section{Results}

We successfully construct high resolution mass models for all five
clusters from the unambiguous galaxy-galaxy lensing signal detected
using the maximum-likelihood analysis. We also detect a cut-off in the
extent of a fiducial halo, which we argue is a result of tidal
stripping of cluster galaxies that traverse through the central
regions of the cluster. All our lens models are plotted in
Figs.~1--5. The maximum-likelihood analysis yields the following: (i)
the mass-to-light ratio in the $V$-band of a typical L$^*$ does
not evolve significantly as a function of redshift, (ii) the fiducial
truncation radius of an $L^*$ varies from about 15 kpc at $z = 0.18$
to 70 kpc at $z = 0.57$, (iii) the typical central velocity dispersion
is roughly 180\,{\rm \,km\,s}$^{-1}$.

We present the luminosity function of early-type confirmed cluster
members chosen as perturbers in all 5 clusters in Fig.~7.  For the
galaxy model (PIEMD) adopted in our analysis, the total mass of an
L$^*$ varies with redshift from $\sim 2.8 \times 10^{11}\,M_{\odot}$
to $\sim 7.7 \times 10^{11}\,M_{\odot}$ The mass-to-light ratios
quoted here take passive evolution of elliptical galaxies into account
as given by the stellar population synthesis models of Bruzual \&
Charlot (2003), therefore any detected trend reflects pure mass
evolution (see Table~2 and Fig.~6). The mass obtained for a typical
bright cluster galaxy by Tyson et al.\ (1998) using only strong
lensing constraints inside the Einstein radius of the cluster
Cl\,0024+1654, at $z = 0.41$, is consistent with our results. All
error bars quoted here are $\sim 3\sigma$. Scaling laws were needed to
relate the mass to the light (see eqn. 30), the effect of the assumed
form on the derived fiducial sub-halo mass of an $L^*$ is show in
Fig.~8.

By construction, the maximum-likelihood technique presented here
provides the mass spectrum of sub-halos in the cluster directly (see
Fig.~8). Note that as stated before in performing the likelihood
analysis to obtain characteristic parameters for the sub-halos in the
cluster it is assumed that light traces mass. This is an assumption
that is well supported by galaxy-galaxy lensing studies in the field
(Wilson et al.\ 2001) as well as in clusters (Clowe \& Schneider
2002). In fact, all lens modeling and rotation curve measurements
suggest an excess of baryons in the inner regions. Note however that
for our choice of mass model (the PIEMD) the mass to light ratio is
not constant with radius within an individual galaxy halo.  Since the
procedure involves a scaled, self-similar mass model that is
parametric, we obtain a mass estimate for the dark halos (sub-halos)
as a function of their luminosity. This provides us with a clump mass
spectrum. Tidal truncation by the cluster causes these galaxy halo
masses to be lower than that of equivalent luminosity field galaxies
at comparable redshifts obtained from galaxy-galaxy lensing. The
fraction of mass in these clumps is only 10--20\% of the total mass of
the cluster within the inner $500\,h^{-1}\,{\rm kpc}$ of these high
central density clusters. The remaining 80--90\% of the cluster mass is
consistent with being smoothly distributed (in clumps with mass
$M\,<\,10^{10}\,M_{\odot}$), the precise composition of this component
depends on the hitherto unknown nature of dark matter. Note that the 
the upper and lower limits on the mass spectrum vary from cluster to
cluster due to the difference in the luminosity functions of cluster
galaxies. These mass functions can now be directly compared to the sub-halo
mass functions of dark matter halos in cosmological N-body simulations,
the results of which are presented elsewhere (Natarajan \& Springel 2004).

\section{Conclusions and Discussion}

In this paper, we present (i) high resolution mass models for lensing
clusters and (ii) the mass function of sub-halos inside these
clusters. Detailed results of the application of our galaxy-galaxy
lensing analysis techniques to five {\it HST} cluster lenses (as well
as a further cluster we have previously analyzed in an identical
manner) are used to construct high resolution mass models of the inner
regions. In order to do so we have utilized both strong and weak
lensing observations for these massive clusters.  The goal is to
quantify substructure in the cluster assuming that the sub-halos
follow the distribution of bright, early-type cluster
galaxies. Similar attempts have been made in the lower density field
environment yielding typical galaxy masses and central velocity
dispersions. The mass distribution for a typical galaxy halo inferred
from field studies are extended with no discernible cut-off. By
contrast in the cluster environments probed in this work we detect an
edge to the mass distribution in cluster galaxies. We have performed
various stringent checks to ascertain that this is not an artifact of
the choice of mass model and rather evidence for tidal stripping by
the global cluster potential well. Aside from the detailed lens
models, we also present the first ever mass spectrum (albeit within a
limited mass range with sub-halo masses ranging from $10^{11} -
10^{12.5}\, \msun$) of substructure in the inner regions of these
clusters. The survival and evolution of substructure offers a
stringent test of structure formation models within the CDM
paradigm. Sub-halos of the scale detected in all these clusters
indicate a high probability of galaxy--galaxy collisions over a Hubble
time within a rich cluster.  However, since the internal velocity
dispersions of these clumps associated with early-type cluster
galaxies ($\sim 150$--250\,{\rm km\,s}$^{-1}$) are much smaller than
their orbital velocities, these interactions are unlikely to lead to
mergers, suggesting that the encounters of the kind simulated in the
galaxy harassment picture by Moore et al.\ (1996) are the most
frequent and likely. High resolution cosmological N-body simulations
of cluster formation and evolution (De Lucia et al.\ 2004; Ghigna et
al.\ 1998; Moore et al.\ 1996), find that the dominant interactions
are between the global cluster tidal field and individual galaxies
after $z = 2$. The cluster tidal field significantly tidally strips
galaxy halos in the inner 0.5 Mpc and the radial extent of the
surviving halos is a strong function of their distance from the
cluster center. Much of this modification is found to occur between $z
= 0.5$--0.  The trends seen in halo size $r_t^*$ with redshift
detected in our analysis of these clusters (Natarajan et al.\ 2002a)
are broadly in agreement with high-resolution cosmological N-body
simulations of currently popular cosmological models (De Lucia et al.\
2004).  Further interpretation and comparison of these results with
theoretical models and high resolution N-body simulations is
presented elsewhere (Natarajan \& Springel 2004).

\acknowledgments

PN acknowledges gratefully support from NASA via HST grant
HST-GO-09722.06-A.  JPK acknowledges support from the CNRS and Caltech,  and 
IRS from the Royal Society and the Leverhulme Trust.

\newpage

%
%
\begin{figure}
\includegraphics{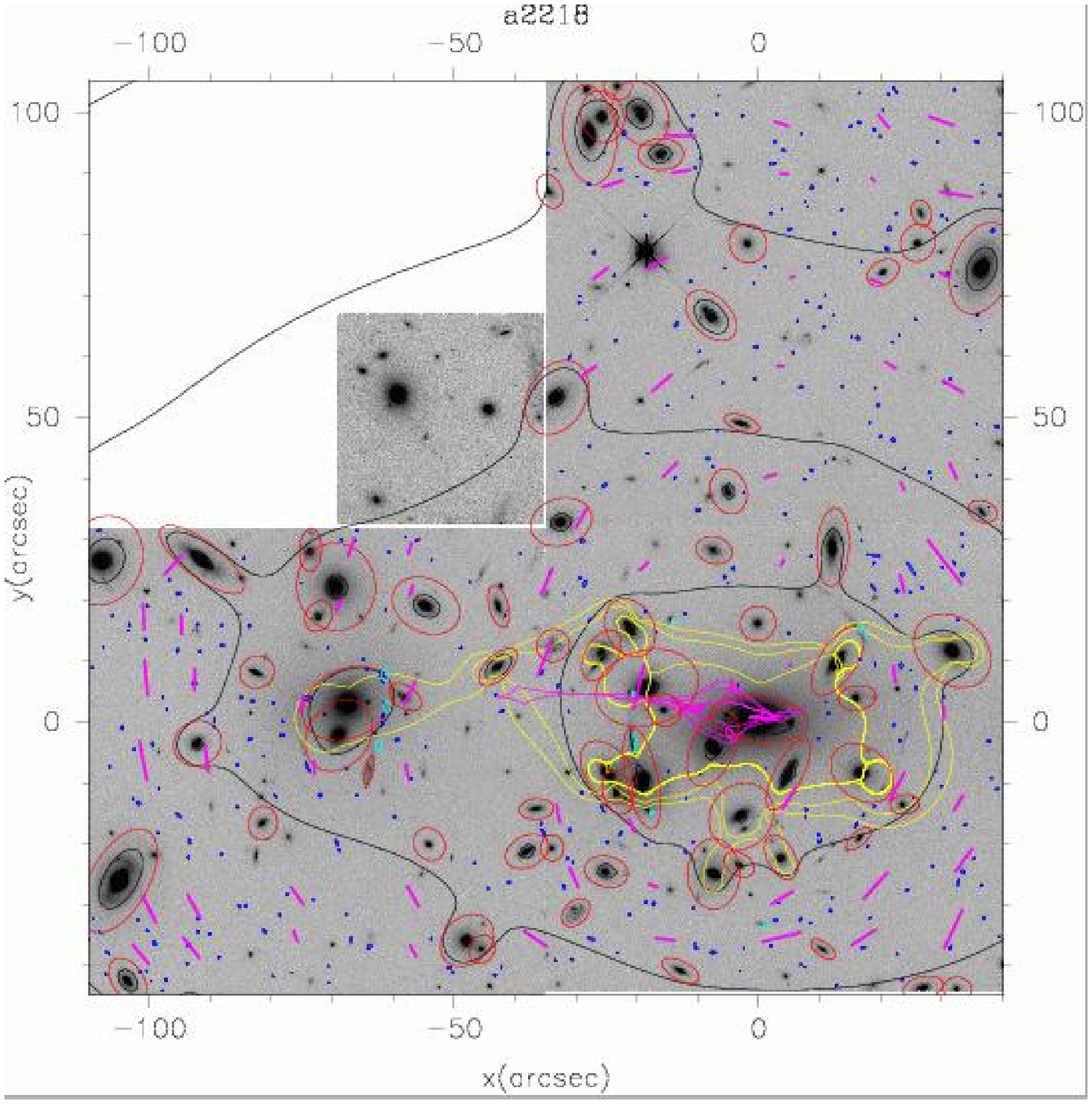}
\caption{The best-fit mass model for A\,2218. The confocal ellipses
are the equi-potentials of the full potential including the
substructure.  The critical curves for the best-fit model computed for
sources at $z = 1, 2 \& 3$ and caustics are also shown. The multiple
images with measured redshifts that constrain the model are shown in
cyan. All cluster members included in the mass modeling are marked
with a red ellipse. The smoothed shear field is shown as pink sticks
and the individual lensed galaxies are marked in blue. The critical
lines and caustics are shown in yellow and pink respectively.}
\end{figure}

%
%
\begin{figure}
\includegraphics{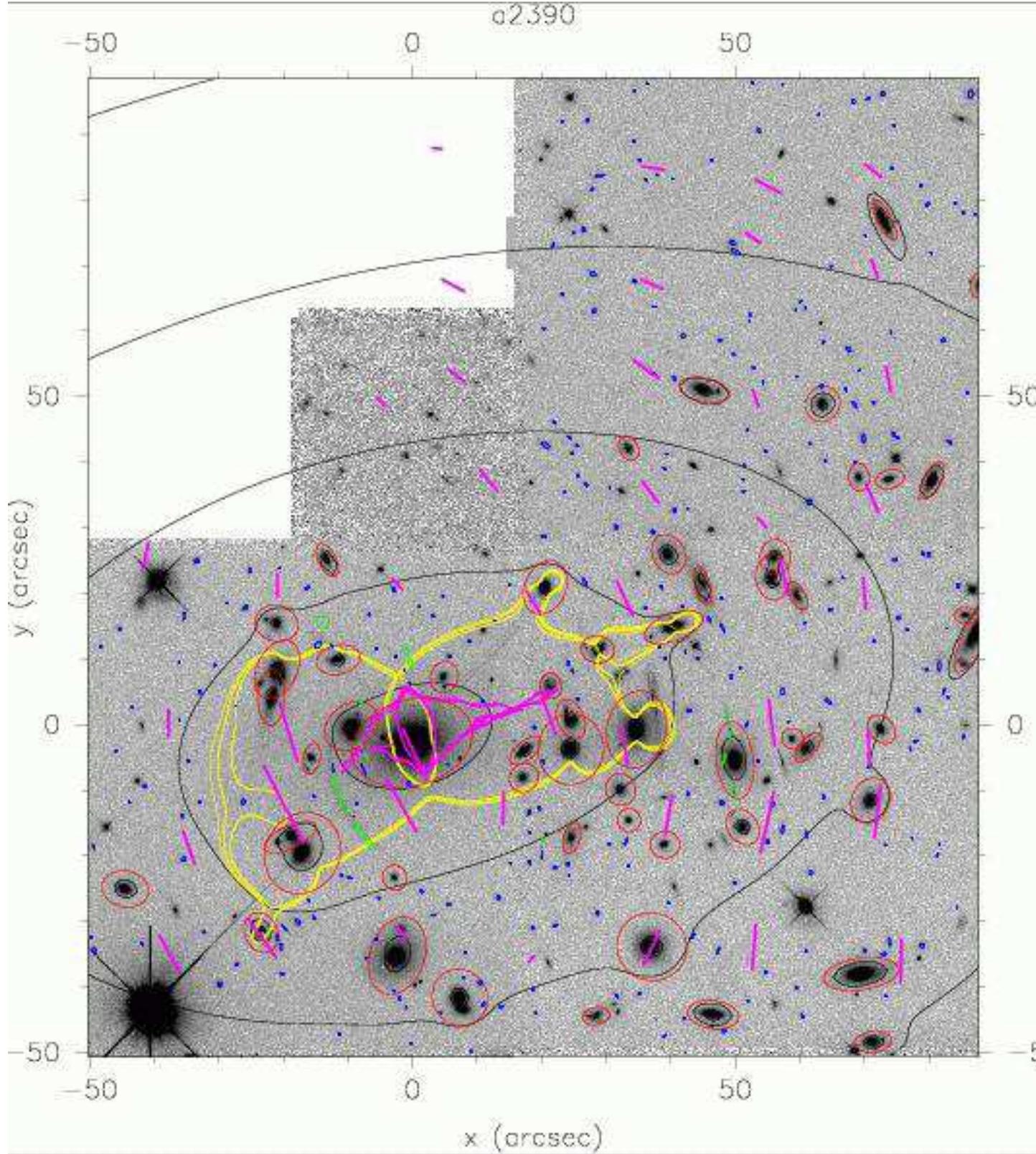}
\caption{Mass model for A\,2390: all included galaxy scale potentials are 
demarcated with ellipses and the critical curves computed for sources
at $z = 1, 2 \& 3$ are also shown here for the best-fit mass model.}
\end{figure}

%
%
\begin{figure}
\includegraphics{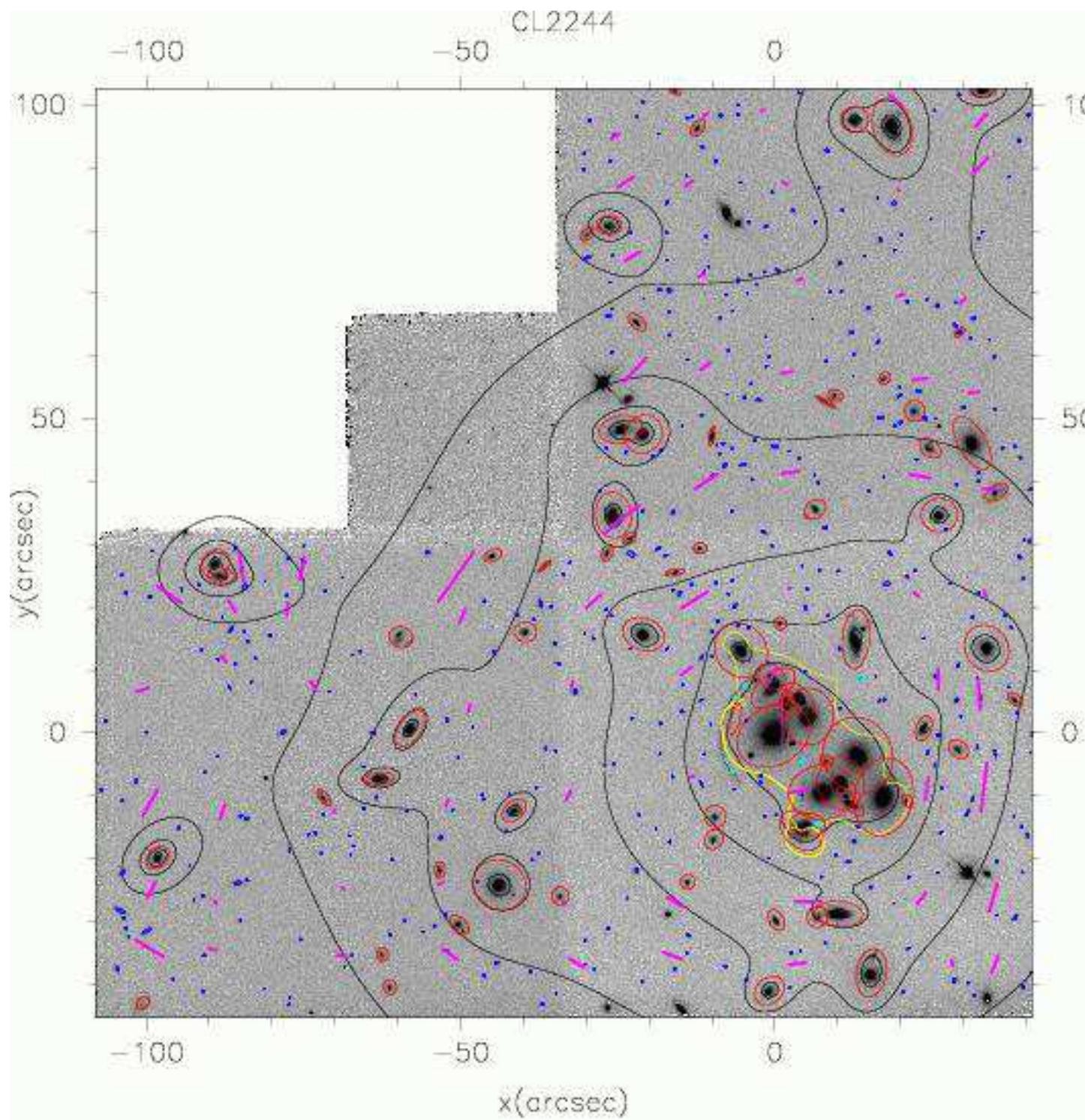}
\caption{Mass model for Cl\,2244.}
\end{figure}

%
%
\begin{figure}
\includegraphics{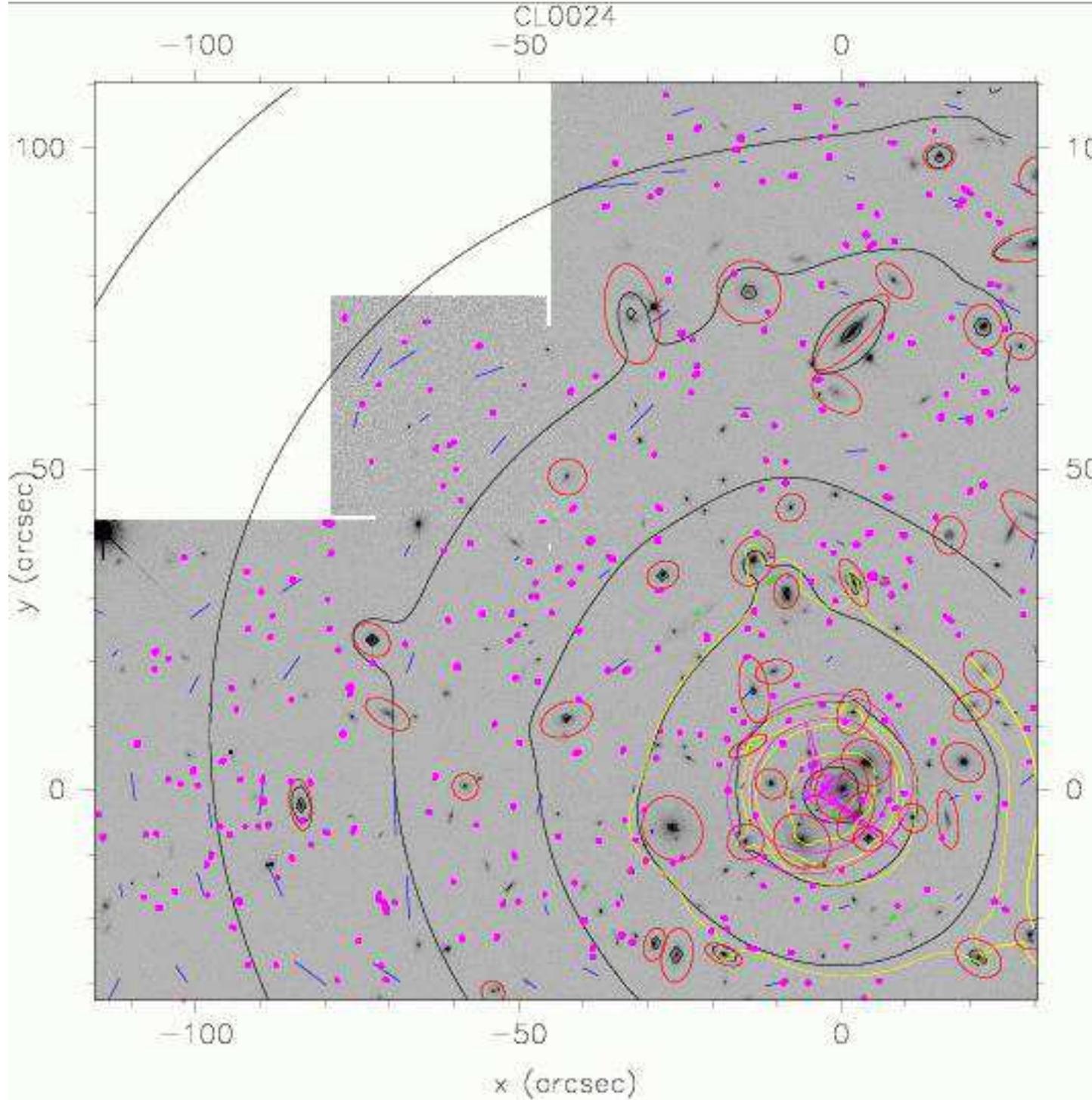}
\caption{Mass model for Cl\,0024+16. Note that this cluster has a 
smooth set of confocal equi-potentials. This model is in good agreement
with the one recently published by Kneib et al.\ (2003) that extends
out to 5 h$^{-1}$ Mpc.}
\end{figure}

%
%
\begin{figure}
\includegraphics{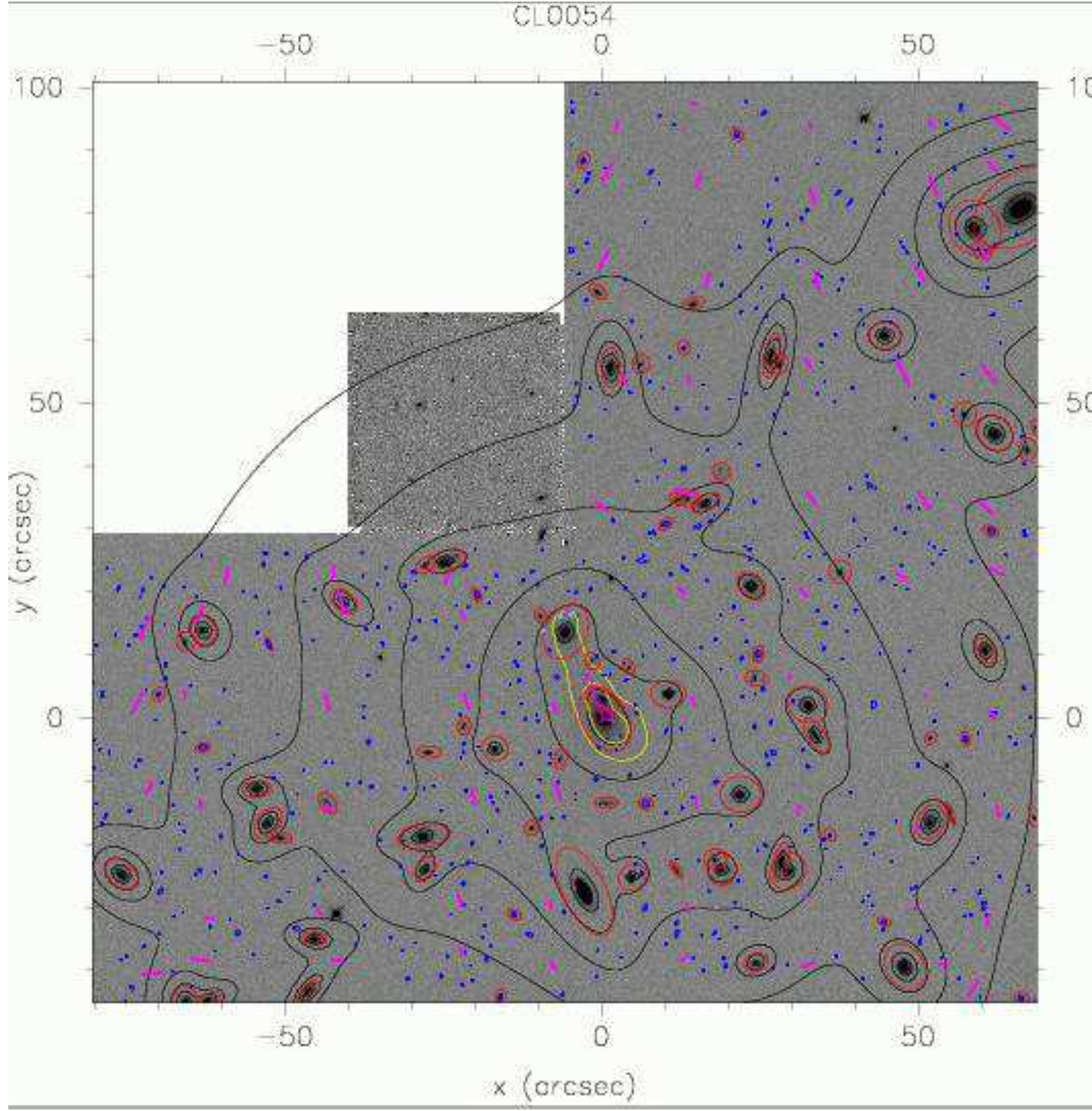}
\caption{Mass model for Cl\,0054$-$27. Although we require 2 smooth
components in the inner most regions, the overall mass model for this
cluster is a lot smoother compared to the others studied here.}
\end{figure}

%
%
\begin{figure}
\begin{center}
\resizebox{13cm}{13cm}{\includegraphics{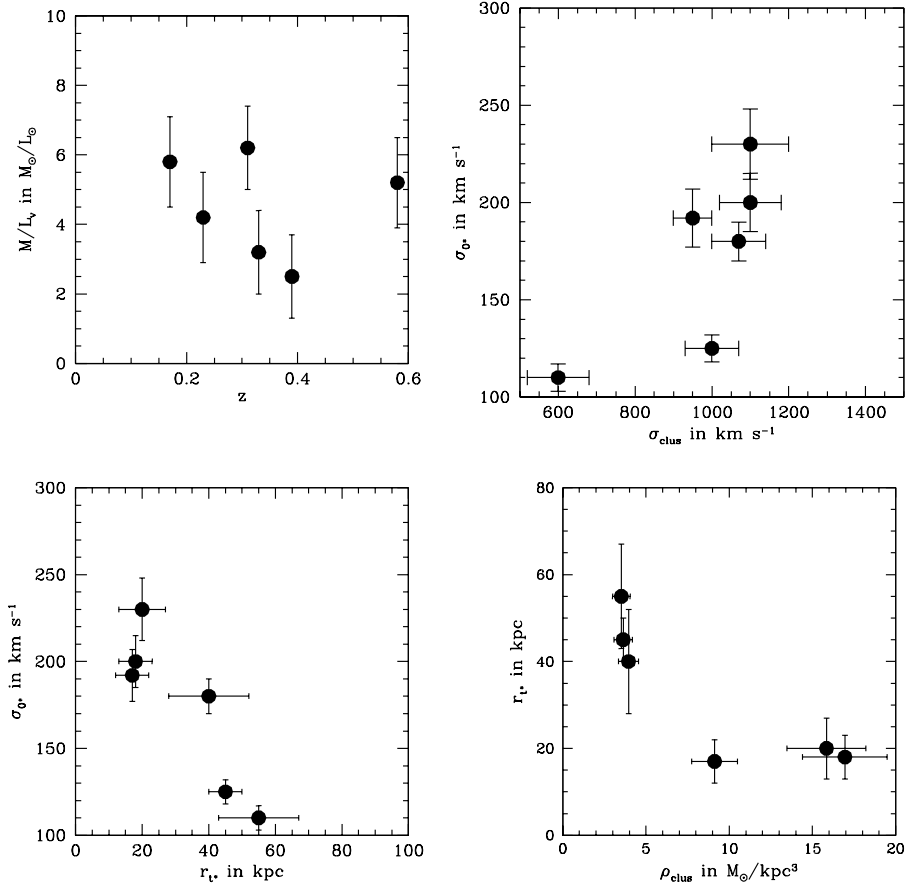}}%
\end{center}
\caption{The results of the maximum likelihood analysis. Top left
  panel: the variation of the mass-to-light ratio in the V-band as
a function of redshift for a typical $L^*$ sub-halo; Top right panel:
  the central velocity dispersion of a typical $L^*$ sub-halo in the
cluster versus the velocity dispersion of the cluster; Bottom left
panel: the central velocity dispersion versus the
tidal radius of an $L^*$ sub-halo which is the result of the maximum
  likelihood analysis; Bottom right panel: the tidal truncation radius
versus central density of the cluster.}
\end{figure}

%
%
\begin{figure}
\begin{center}
\resizebox{13cm}{13cm}{\includegraphics{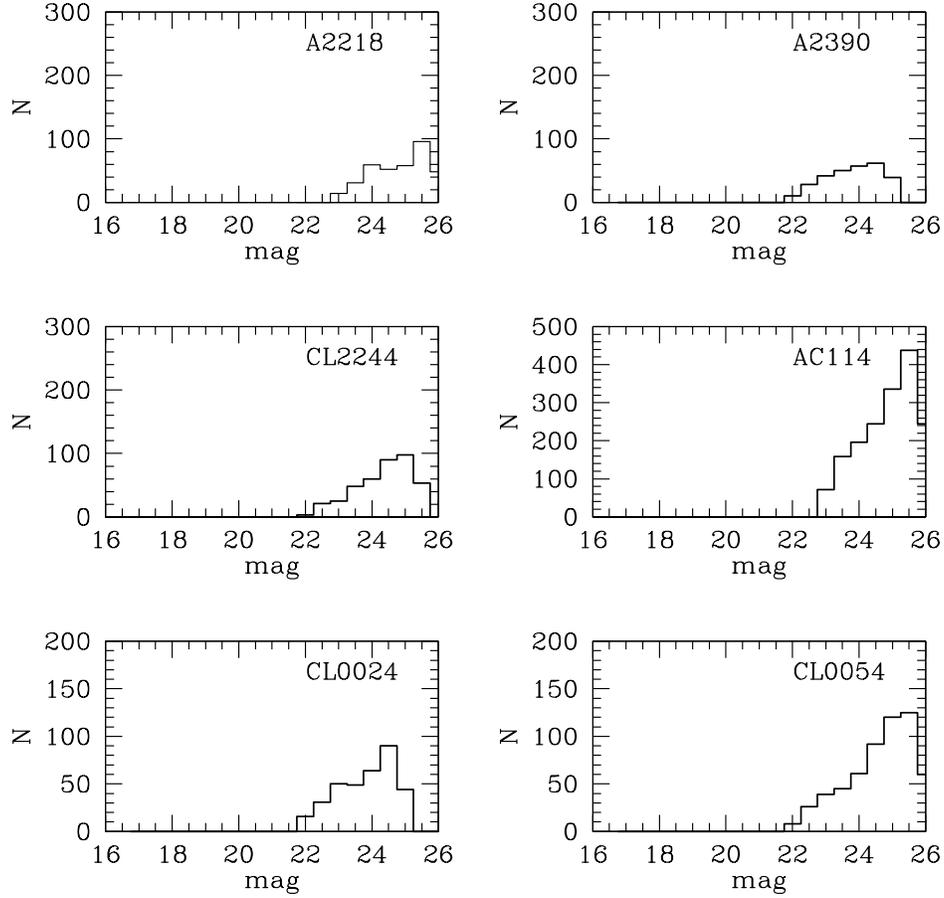}}%
\end{center}
\caption{Luminosity function of the selected early-type galaxies. For
four clusters (A\,2218, AC114, Cl\,0024+16, and Cl\,0054-27)
morphological classification and membership information was obtained
from the MORPHS collaboration. For A\,2390 and Cl\,2244-02 a
magnitude cut was used to delineate cluster members.}
\end{figure}

%
%
\begin{figure}
\begin{center}
\resizebox{13cm}{7.5cm}{\includegraphics{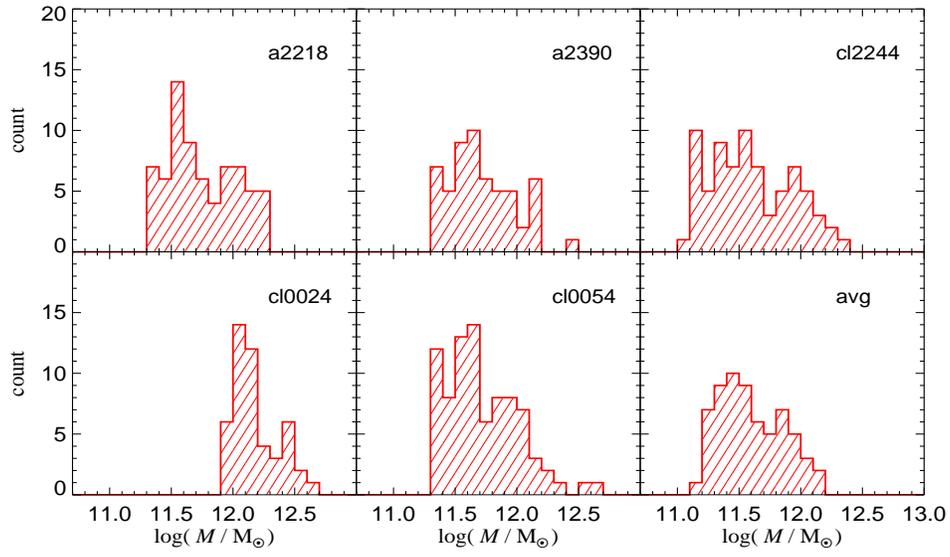}}%
\end{center}
\caption{Sub-halo mass spectrum retrieved from the maximum-likelihood
analysis for the five {\it HST} cluster-lenses studied here.}
\end{figure}

%
%
\begin{figure}
\begin{center}
\resizebox{13cm}{13cm}{\includegraphics{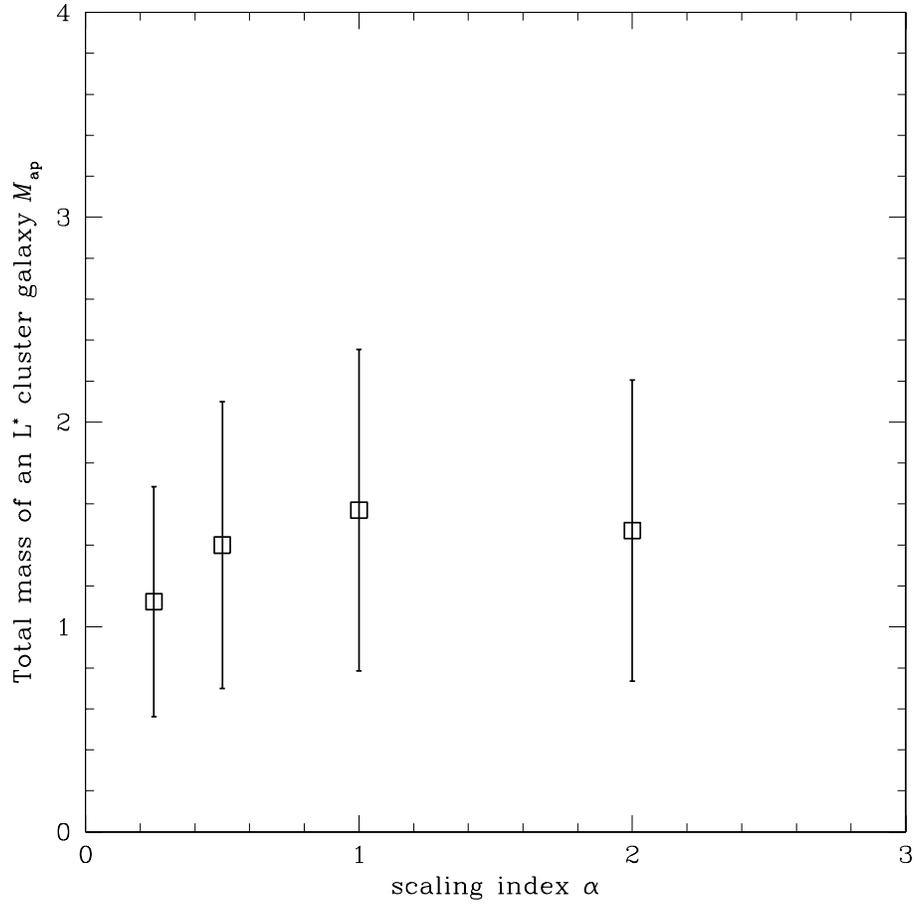}}%
\end{center}
\caption{The retrieved mass $M_{\rm ap}$ for an $L^*$ cluster galaxy
in A2218 as a function of the choice of the scaling parameter $\alpha$
that tunes the outer edge of the mass distribution used to model the
sub-halos.}
\end{figure}

\end{document}